\documentclass[12pt,preprint]{aastex}
\usepackage{emulateapj5,epsf}

\makeatletter
\newcounter{subeqncnt}
\def\thesubeqncnt{\alph{subeqncnt}}
\def\subequations{\begingroup%
\stepcounter{equation}\edef\@tempa{\theequation}%
\let\c@equation\c@subeqncnt\c@subeqncnt\z@
\edef\theequation{\@tempa\noexpand\thesubeqncnt}}

\makeatother

\slugcomment{accepted ApJ, October 14, 2009}

\shorttitle{
High-$z$ QSO formation
}
\shortauthors{Kawakatu and Wada}

\begin{document}

\title{
Formation of High-redshift ($z>6$) Quasars Driven by Nuclear Starbursts
}


\author{Nozomu Kawakatu\altaffilmark{1,2} and 
Keiichi Wada\altaffilmark{1,3}}

\altaffiltext{1}{National Astronomical Observatory of Japan, 2-21-1 Osawa, 
Mitaka, Tokyo 181-8588, Japan; kawakatu@th.nao.ac.jp}
\altaffiltext{2}{JSPS fellow}
\altaffiltext{3}{Graduate School of Science and Engineering, Kagoshima University, Kagoshima 890-8580, Japan; wada@cfca.jp}


\begin{abstract}
Based on the physical model of a supermassive black hole (SMBH)
growth via gas accretion in a circumnuclear disk (CND) proposed by 
Kawakatu \& Wada (2008), we describe the formation of high-$z$ ($z > 6$) 
quasars (QSOs) whose BH masses are $ M_{\rm BH}> 10^{9}M_{\odot}$. 
We derive the necessary conditions to form QSOs at $z > 6$ by only gas 
accretion: 
(i) A large mass supply with $M_{\rm sup} > 10^{10}M_{\odot}$ 
from host galaxies to CNDs, because the final BH mass is 
only $1-10\%$ of the total supplied mass from QSO hosts. 
(ii) High star formation efficiency for a rapid BH growth 
which is comparable to high-$z$ starburst galaxies such as submillimeter 
galaxies (SMGs). 
We also find that if the BH growth is limited by the Eddington accretion, 
the final BH mass is greatly suppressed when the period of mass-supply from 
hosts, $t_{\rm sup}$ is shorter than the Eddington timescale. 
Thus, the super-Eddington growth is required for the QSO formation 
as far as $t_{\rm sup}$, which is determined by the efficiency of angular momentum transfer, is shorter than $\sim 10^{8}\,{\rm yr}$. 
The evolution of the QSO luminosity depends on the redshift $z_{\rm i}$ 
at which accretion onto a seed BH is initiated. 
In other words, the brighter QSOs at $z >6 $ favor the late growth of SMBHs 
(i.e., $z_{\rm i}\approx 10$) rather than early growth (i.e., $z_{\rm i}
\approx 30$). For $z_{\rm i}\approx 10$, $t_{\rm sup}\simeq 10^{8}\,{\rm yr}$ 
is shorter than that of the star formation in the CND. Thus, the gas in the CND can accrete onto a BH more efficiently, compared with the case for $z_{\rm i}\approx 30$ (or $t_{\rm sup}\approx 10^{9}\,{\rm yr}$). 
Moreover, we predict the observable properties and 
the evolution of QSOs at $z >6$. In a QSO phase, there should exist a 
stellar rich massive CND, whose gas mass is about $10\%$ of the dynamical 
mass inside $\sim 0.1-1\,{\rm kpc}$. 
On the other hand, in a phase where the BH grows (i.e., a proto-QSO phase), 
the proto-QSO has a gas rich massive CNDs whose gas mass is comparable to the 
dynamical mass. 
Compared with the observed properties of the distant QSO SDSS J1148+5251 
observed at $z=6.42$, we predict that SDSS J1148+5251 corresponds to 
the scenario of the late growth of SMBH with $z_{\rm i}\sim10$, which is 
accompanied by a massive CNDs with $M_{\rm g}\approx 
5\times 10^{10}M_{\odot} $ and the luminous nuclear starburst $L_{\rm SB}$ 
at infrared band with $L_{\rm SB}\approx 10^{47}\,{\rm erg}\,{\rm s}^{-1}$. 
Moreover, we predict that the progenitor of SDSS J1148+5251 can be the 
super-Eddington object. These predictions can be verified by ALMA, SPICA 
and JWST. 
\end{abstract}

\section{Introduction}
Supermassive black holes (SMBHs) with masses in the range 
of $10^{6}-10^{9}M_{\odot}$ are the engines that power 
active galactic nuclei (AGNs) and quasars (QSOs). 
There is also ample evidence that SMBHs reside at the center 
of most galaxies (e.g., Kormendy \& Richstone 1995; Richstone et al. 1998; 
Ho et al. 1999), including the Milky Way (e.g., Genzel et al. 1997; 
Sch{\"o}del et al. 2002; Ghez et al. 2003). 
In the local Universe, there are tight correlations between the masses of 
SMBHs and the masses and velocity dispersions of the spheroidal components 
(bulge) of the hosts (e.g., Kormendy \& Richstone 1995; Richstone et al. 1998; 
Ferrarese \& Merritt 2000; Gebhardt et al. 2000; Tremaine et al. 2002; 
Marconi \& Hunt 2003; H{\"a}ring \& Rix 2004; Barth et al. 2005). 
These observational facts suggest that an intriguing link between the formation of bulges and SMBHs. 

High redshift QSOs are essential to understand the formation 
and evolution of SMBHs. 
In recent years, more than thirty QSOs at $z\approx 6$ 
have been discovered (Fan et al. 2000, 2001, 2003, 2004, 2006; Goto 2006; 
Jiang et al. 2006; Willott et al. 2007; Jiang et al. 2009).  
The higher luminosity of the distant QSO at $z=6.42$, corresponding to SDSS J1148+3251 (Fan et al. 2003) with $\sim 10^{47}\,{\rm erg}\,{\rm s}^{-1}$, 
implies that a SMBH with mass $\geq 10^{9}M_{\odot}$ is already 
in place within $\sim$ 1 Gyr 
after the big bang, by assuming the Eddington luminosity. 
These requirements set significant constraints on the evolution 
and formation of SMBHs in the early Universe. 
There are some analytical and semi-analytical studies 
(Haiman \& Loeb 2001; Haiman 2004; Yoo \& Mirald-Escud\'{e} 2004; Shapiro 2005; Volonteri \& Rees 2006; Tanaka \& Haiman 2009) and numerical simulations 
(Li et al. 2007; Sijacki et al. 2009) that discuss the formation of 
$> 10^{9}M_{\odot}$ SMBH at $z >6$.
A key physical process governing evolution of QSOs is the mass accretion 
toward a SMBH, although BH growth via merging may be expected at the high 
mass BH with $M_{\rm BH} > 10^{9}M_{\odot}$ (e.g., 
Shankar 2009 and references therein). 
Although the accretion process is a complicated phenomenon 
over nine orders of magnitude in the size scale, the Eddington-limited 
accretion rate was simply assumed in most previous studies. 
However, the evolution of the SMBH is not only controlled by the accretion 
processes in the vicinity of the SMBH, but also related to the mass supply 
from the host galaxy to the circumnuclear region.

A number of mechanisms for gas accumulation on the $\sim 1-10$ kpc scale 
down to the galactic central region have been proposed, 
e.g., the tidal torque driven by the major and minor mergers of galaxies 
(e.g., Toomre \& Toomre 1972; Mihos \& Hernquist 1994, 1996; Saitoh \& Wada 
2004; Saitoh et al. 2009) and the stellar bars (e.g., Noguchi 1988; 
Shlosman et al. 1990; Fukuda et al. 1998; Fukuda et al. 2000; 
Maciejewski et al. 2002; Namekata et al. 2009), gas drag and dynamical 
friction in the dense stellar 
cluster (Norman \& Scoville 1988) and the radiation drag (e.g., 
Umemura et al. 1997; Umemura 2001; Kawakatu \& Umemura 2002). 
However, the accumulated gas does not directly accrete onto a SMBH, 
since the angular momentum of the gaseous matter cannot be thoroughly removed. 
Thus, some residual angular momentum would terminate the radial infall, 
so the accreted gas forms a reservoir, 
i.e., a circumnuclear disk (CND), in the central $\sim$100 pc around a SMBH 
whose scale depends on the angular momentum of the gas. 
If the gravitational instability takes place in a CND, star formation is 
naturally expected in this region. 
Active nuclear star formation ($< 100\,{\rm pc}$) has been actually 
observed in nearby AGNs (e.g., Imanishi \& Wada 2004; Davies et al. 2007; 
Watabe et al. 2008; Chen et al. 2009; Hicks et al. 2009). 
It is theorized that the nuclear starburst could obscure 
some types of AGNs (e.g., Ohsuga \& Umemura 1999; Wada \& Norman 2002: hereafter WN02; Thompson et al. 2005; Watabe et al. 2005; 
Ballantyne 2008; Schartmann et al. 2008; Wada et al. 2009). 
The nuclear starburst also affects the growth of SMBHs, 
because the radiation and/or supernova feedback from starbursts 
can enhance the mass accretion onto a SMBH (e.g., Norman \& Scoville 1988; 
Umemura et al. 1997; WN02; Vollmer \& Beckert 2003; Vollmer, Beckert \& Davis 
2008; Collin \& Zahn 2008). 
In order to reveal the final rate of mass accretion to the BH region,  
it is crucial to {\it link mass accretion processes from a galactic scale 
with those from an accretion disk in the vicinity of a central BH, 
via the CND}. Recently, we have proposed a theoretical model of 
a nuclear starburst disk supported by the turbulent pressure led 
by supernova explosions (Kawakatu \& Wada 2008: hereafter KW08). 
In KW08, the turbulence excited by supernovae transports the angular momentum. 
We showed how a SMBH grows from a seed BH, taking into account the mutual connection between the mass-supply from a host galaxy and the physical states of the CND accompanied by the star formation. 
This theoretical model should be confirmed for the formation of high-$z$ 
($z >6$) QSOs whose masses are $\simeq 10^{9}M_{\odot}$.

This paper is organized as follows. 
We briefly outline the  model in KW08 in $\S 2$. 
By adopting KW08, we will show the necessary conditions to form QSOs 
at $z > 6$ in terms of the total accreted gas mass from host galaxies 
and star formation efficiency in the CNDs ($\S 3$). 
In addition, we will predict observable properties and the evolution of QSOs 
at $z >6$ ($\S4$). 
In section 5, we discuss how the distant QSO J1148+5251 at $z=6.42$ 
forms and predict the nature of the early phase of this QSO. 
Section 6 is devoted to our summary. 
Unless otherwise stated, all results shown below refer to the currently 
favored $\Lambda$ cold dark matter model with $\Omega_{\rm M}=0.24$, 
$\Omega_{\Lambda}=0.76$, $h=0.73$, $\Omega_{\rm b}=0.042$, 
$\sigma_{8}=0.74$ and $n=0.95$ (Spergel et al. 2007).

\section{Models}
We here briefly describe a coevolution model of SMBHs and 
CNDs (KW08). We assume that dusty gas 
is supplied to a region around a central SMBH 
at a constant rate of $\dot{M}_{\rm sup}$ from a host galaxy 
whose surface density $\Sigma_{\rm host}$, 
including the gas and stellar components (Fig. 1 in KW08). 
The accumulated gas forms a clumpy CND around a central SMBH with $M_{\rm BH}$,  which is vertically supported by turbulent pressure via SN explosions (WN02; Vollmer \& Beckert 2003; Vollmer et al. 2008; Collin \& Zahn 2008; 
Wada et al. 2009). 
We here assume the isothermal cold gas dominates mass 
($T_{\rm g}=50-100\,{\rm K}$) in the CND since the molecular and 
dust cooling is effective (e.g., Wada \& Tomisaka 2005; Wada et al. 2009). 

\subsection{Turbulent pressure-supported CND}
On the vertical structure of the CND, we assume that 
the turbulent pressure associated with SN explosions is balanced to gravity, 
$g$ caused by 
\begin{equation}
\rho_{\rm g}(r)v_{\rm t}^{2}(r) = \rho_{\rm g}(r)g(r)h(r), 
\end{equation}
where $\rho_{\rm g}(r)$, $v_{\rm t}(r)$ and $h(r)$ are the gas density, 
the turbulent velocity and the scale height of the disk, respectively. 
Here, the gravity, $g(r)$ is obtained as 
$g(r)\equiv GM_{\rm BH}h/r^{3}+\pi G(\Sigma_{\rm disk}(r)+\Sigma_{\rm host})$ 
where $\Sigma_{\rm disk}(r)$ is the surface density of baryonic components 
(the gaseous matter and stars). 
The geometrical thickness is determined by the balance between the 
turbulent energy dissipation and the energy input from SN explosions as follows.
\begin{equation}
\frac{\rho_{\rm g}(r)v_{\rm t}^{2}(r)}{t_{\rm dis}(r)}
=\frac{\rho_{\rm g}(r)v_{\rm t}^{3}(r)}{h(r)}
= \eta_{\rm SN} S_{*}(r)E_{\rm SN},
\end{equation}
where the dissipation timescale of the turbulence, $t_{\rm dis}(r)
=h(r)/v_{\rm t}(r)$, $E_{\rm SN}$ is the total energy ($10^{51}\,{\rm erg}$) 
injected by an SN and $\eta_{\rm SN}$ is heating efficiency per unit mass which denotes how much energy from SNe is converted to kinetic energy of the matter. 
Recent observational studies on the relationship between
star formation rate and gas surface density in nearby galaxies
suggest that the star formation rate is approximately proportional to 
the gas density for high density region ($n_{\rm H} > 10^{2}\,{\rm cm}^{-3}$), 
i.e., $S_{*}=C_{*}\rho_{\rm g}^{n}$ with $n\sim 1$ (e.g., Bigiel et al. (2008)). Here $C_{*}$ is the star formation efficiency. 
Theoretical studies based on numerical simulations of the inter stellar 
medium (ISM) also support this proportionality for the high-density end 
(Wada \& Norman 2007; Dobbs \& Pringle 2009; Krumholz et al. 2009). 
Since we are interested in the evolution of dense CNDs with 
$n_{\rm H} > 10^{2}\,{\rm cm}^{-3}$, we here suppose $S_{*}=C_{*}\rho_{\rm g}$.

The star formation time scale or $C_{*}^{-1}$ is of the order of an orbital 
period in a quasi-steady, self-regulated galactic disk, in which the Toomre
Q-value is roughly unity. However, observations suggest that the
local star formation rate for a given gas density varies widely
distributed by one to two orders of magnitude (Komugi et al. 2005; 
Bigiel et al. 2008, see also Fig. 5 and discussion in $\S 3.1$). 
Based on a theoretical model of the ISM and star formation, 
Wada \& Norman (2007) claimed that this wide variety could be
caused by a wide range of star formation efficiency. 
In general, the star formation rate (or efficiency) on small scale is not simply determined only by gas density (see e.g. Kawamura et al. 2009 
for molecular clouds in LMC).  The physical background of this variety 
is still unclear even in our Galaxy. This is also the case in the galactic 
central region. Although many local spiral galaxies show molecular gas 
concentration in the central sub-kpc region (e.g. Komugi et al. 2008), 
the star forming activity is not uniquely determined by the gas density. 
In fact, the local star formation rate is correlated with a fraction of 
high density molecular gas (Muraoka et al. 2009). Moreover, at high redshift,  
the star formation efficiency in the galactic central region is hardly determined observationally and theoretically. 
The circumnuclear gas disk is not necessarily in a self-regulated state, 
since the rotational time scale is comparable to 
the life time of massive stars or smaller. 
Therefore under frequent mergers of small gas-rich galaxies at high redshift, 
which would be a main cause of mass supply toward a galactic center, 
the circum-nuclear region might be far from in an equilibrium. 
Considering all these uncertainties, we here assume that the star formation
time scale is a free parameter, and examine the results for a wide range of
$C_{*}$.

By using equations (1) and (2), $v_{\rm t}(r)$ and $h(r)$ can be obtained 
(see also eqs. (4) and (5) in KW08). 
For given $r$ and $M_{\rm BH}$, it is found that $v_{\rm t}\propto 
C_{*}^{1/2}$ and $h\propto C_{*}^{1/2}$ (see also WN02). 
The velocity dispersion of molecular hydrogen in CNDs is in fact 
positively correlated with the star formation rate (Hicks et al. 2009).

\subsection{Two modes of gas accretion in CND}
We suppose a kinetic viscosity as a source of angular momentum transfer 
in the gas disk. Then, the mass accretion rate in a viscous accretion disk 
is given by 
\begin{equation}
\dot{M}(r)=2\pi\nu_{\rm t}(r)\Sigma_{\rm g}(r)\left|\frac{d\,{\rm ln}\,\Omega(r)}{d\,{\rm ln}\, r}\right|, 
\end{equation}
where the viscous parameter is $\nu_{\rm t}(r)=\alpha v_{\rm t}(r)h(r)$, 
$\Sigma_{\rm g}$ is the surface density of the gas component in the CND
 and the angular velocity $\Omega(r)$ is given by the radial centrifugal 
 balance. Here $\alpha (\leq 1)$ is a constant in time. 
 The inner radius of the CND ($r_{\rm in}$) is determined by the dust 
sublimation radius, i.e., $r_{\rm in}=3\,{\rm pc}\,(M_{\rm BH}/10^{8}
M_{\odot})^{1/2}$ (see KW08 in details). 
At $r_{\rm in}$ we assume the CND connects with the steady accretion disk.
With respect to the stability, we adopt Toomre's stability criterion, 
i,e., when the surface density of gas in the CND, $\Sigma_{\rm g}$ 
is higher (lower) than the critical surface density, $\Sigma_{\rm crit}$ 
the CND is gravitationally unstable (stable). 
The critical surface density is obtained as 
$\Sigma_{\rm crit}(r)=\kappa(r) c_{\rm s}/\pi G$, 
where $\kappa(r) \equiv 4\Omega^{2}(r)+2\Omega(r) rd\Omega(r)/dr$ 
is the epicyclic frequency and $c_{\rm s}$ is the sound velocity. 
The critical radius $r_{\rm c}$ is determined by the Toomre criterion, 
that is, $\Sigma_{\rm g}(r_{\rm c})=\Sigma_{\rm crit}(r_{\rm c})$. 
In this picture, we naturally expect two modes of gas accretion 
rate as follows: {\it mode (i):} If $r_{\rm c} < r_{\rm in}$ (i.e., the CND is 
fully gravitationally unstable), then the CND is geometrically thick due to 
SN energy feedback, and as a result we have a large accretion rate. We here suppose $\alpha_{\rm SN}=1$ 
motivated by numerical simulations demonstrated by WN02. {\it mode (ii):} If $r_{\rm c} > r_{\rm in}$, 
the scale height of the inner region would be much smaller than mode (i), 
because the scale height is determined by the thermal pressure, 
$P_{\rm g}(r)=\rho_{\rm g}(r)gh(r)$, where $P_{\rm g}(r)=
\rho_{\rm g}(r)c_{\rm s}^{2}$. Here $c_{\rm s}=(5kT_{\rm g}/3m_{\rm p})^{1/2}$, where $k$ and $m_{\rm p}$ are the Boltzman constant and the proton mass. 
In mode (ii), the magneto-rotational instability could be a source of 
turbulence, but the turbulent velocity is comparable to or even smaller than 
the sound speed (e.g., Balbus \& Hawley 1991; Machida et al. 2000). 
As a result, the accretion is less efficient than the mode (i).
We assume $\alpha_{\rm MRI}=0.5$ and $v_{\rm t}=c_{\rm s}$ in mode (ii). 
Note that the adopted $\alpha_{\rm MRI}=0.5$ is relatively larger than that 
derived by numerical simulations (e.g., Machida et al. 2000), but we confirmed 
that the following results do not change at all, even if we use smaller 
$\alpha_{\rm MRI}$ (e.g., $\alpha_{\rm MRI}=0.01$).

\subsection{Coevolution of SMBHs and CNDs}
Our main purpose is to evaluate growth of SMBH, 
the star formation rate and gas mass in the CND. 
We here focus on the time dependence of characteristic radius in the disk 
($r_{\rm c}$, $r_{\rm in}$, and $r_{\rm out}$), instead of solving the 
evolution of radial structure of CND. The radial surface density distribution 
of CND is assumed to be $\Sigma_{\rm disk}(r)=\Sigma_{\rm disk,0}(r/r_{\rm out})^{-\gamma}$ where $r_{\rm out}$ is the outer radius of the CND. 
Hereafter, we assume $\gamma=1$ although the dependence of $\gamma$ on 
$\dot{M}_{\rm BH}$ is weak  for $0 < \gamma < 2 $ (see $\S 3.2$ in KW08). 
In this model the gas supplied from the host galaxy 
is eventually consumed to form the SMBH and stars.  
Thus, the time-evolution of the gas mass in the disk, 
$M_{\rm g}\equiv\int^{r_{\rm out}}_{r_{\rm in}} 2\pi r^{\prime}
\Sigma_{\rm g}(r^{\prime})dr^{\prime}$, is simply given by the 
mass conservation: 
\begin{equation}
M_{\rm g}(t)=\int^{t}_{0}[\dot{M}_{\rm sup}(t^{\prime})
-\dot{M}_{*}(t^{\prime})
-\dot{M}_{\rm BH}(t^{\prime})]dt^{\prime}, 
\end{equation}
where $\dot{M}_{\rm sup}(t)$, $\dot{M}_{*}(t)$ 
and $\dot{M}_{\rm BH}(t)$ are the mass-supply rate from hosts, 
the star formation rate, and the growth rate of SMBH, respectively. 
Here we ignore the mass loss from stars and from CNDs due to 
the starburst wind. 
The time-evolution of SMBH mass $M_{\rm BH}(t)$ 
is obtained as 
$M_{\rm BH}(t)=M_{\rm BH,seed}+\int^{t}_{0}\dot{M}_{\rm BH}
(t^{\prime})dt^{\prime}$, 
where we assume the mass of seed BHs, $M_{\rm BH,seed}=10^{2}M_{\odot}$, 
as end-products of the first generation stars (e.g., Fryer et al. 2001; 
Heger et al 2003; Omukai \& Palla 2003; Yoshida et al. 2006). 
For $\dot{M}_{\rm sup}$, we can assume any function for the mass supply rate, but here we simply take a step function as the first attempt as 
$\dot{M}_{\rm sup}(t)=const$ for $t \leq t_{\rm sup}$, while $\dot{M}_{\rm sup}(t)=0$ for $t > t_{\rm sup}$ where $t_{\rm sup}$ is a period of the mass-supply 
from host galaxies. Thus, key parameters of the SMBH growth and the state of 
CNDs are $\dot{M}_{\rm sup}$, $t_{\rm sup}$ and $C_{*}$.

The growth rate of SMBH, i.e., $\dot{M}_{\rm BH}$, is not necessarily 
equal to the mass accretion rate at the inner boundary, $\dot{M}(r_{\rm in})$. 
The growth rate could be limited by the Eddington accretion rate, 
$\dot{M}_{\rm Edd}$ where $\dot{M}_{\rm Edd}\equiv L_{\rm Edd}/c^{2}$ 
and $L_{\rm Edd}=4\pi cGM_{\rm BH}m_{\rm p}/\sigma_{\rm T}$ with 
$\sigma_{\rm T}$ being the Thomson cross section. 
However, based on radiation hydrodynamic simulations, 
it is also claimed that $\dot{M}_{\rm BH} > \dot{M}_{\rm Edd}$, 
i.e. super-Eddington accretion, could be possible since 
both the radiation field and mass accretion flow around a SMBH
is non-spherical (Ohusga et al. 2005, 2007; Ohsuga \& Mineshige 2007).
Thus we here consider two possible models for $\dot{M}_{\rm BH}$.

\begin{eqnarray}
{\rm (i)} & &\dot{M}_{\rm BH}=\dot{M}(r_{\rm in}). 
\end{eqnarray}
The BH grows with a super Eddington rate 
if $\dot{M}(r_{\rm in}) > \dot{M}_{\rm Edd}$.

\begin{eqnarray}
{\rm (ii)} & &\dot{M}_{\rm BH}=\epsilon_{\rm BH}^{-1}\dot{M}_{\rm Edd},
\end{eqnarray}
where $\epsilon_{\rm BH}=0.1$ is the energy conversion efficiency.
We here assume that there is a mass ``outflow'' with a rate,  
$\dot{M}_{\rm outflow}=\dot{M}(r_{\rm in})-\dot{M}_{\rm BH}$. 
Hereafter we call the model (i) and model (ii) as the super-Eddington growth 
model and the Eddington-limited growth one, respectively. 
We first show results of the super-Eddington accretion in $\S$$\S$3.1 
and 4.1. 
Results based on the Eddington-limited accretion are discussed on 
$\S$$\S$3.2 and 4.2.

Finally, we should stress what happens if we use the non-linear Kennicutt
-Schmidt's law, i.e., $S_{*}=C_{*}^{\prime}t_{\rm ff}^{-1}\rho_{\rm g}^{1.5}$, 
where $t_{\rm ff}$ is the free-fall time scale. 
It is found that the BH growth rate is then described by 
$\dot{M}_{\rm BH}\propto 
C_{*}^{\prime}\rho_{\rm g}\propto M_{\rm g}$ because of the turbulent 
viscosity $\nu_{\rm t}\propto C_{*}^{\prime}$, while the star formation rate 
scales as $\dot{M}_{*}\propto M_{\rm g}^{3/2}$. 
Thus, the gas supplied from host galaxies turns into stars in the CND 
more efficiently, compared with the case for $S_{*}\propto \rho_{\rm g}$ 
(see eq. (4)). It implies that the final BH mass becomes smaller  
due to the efficient star formation, if we adopt non-linear 
Kennicutt-Schmidt's law. 

\vspace{5mm}
\epsfxsize=8cm 
\epsfbox{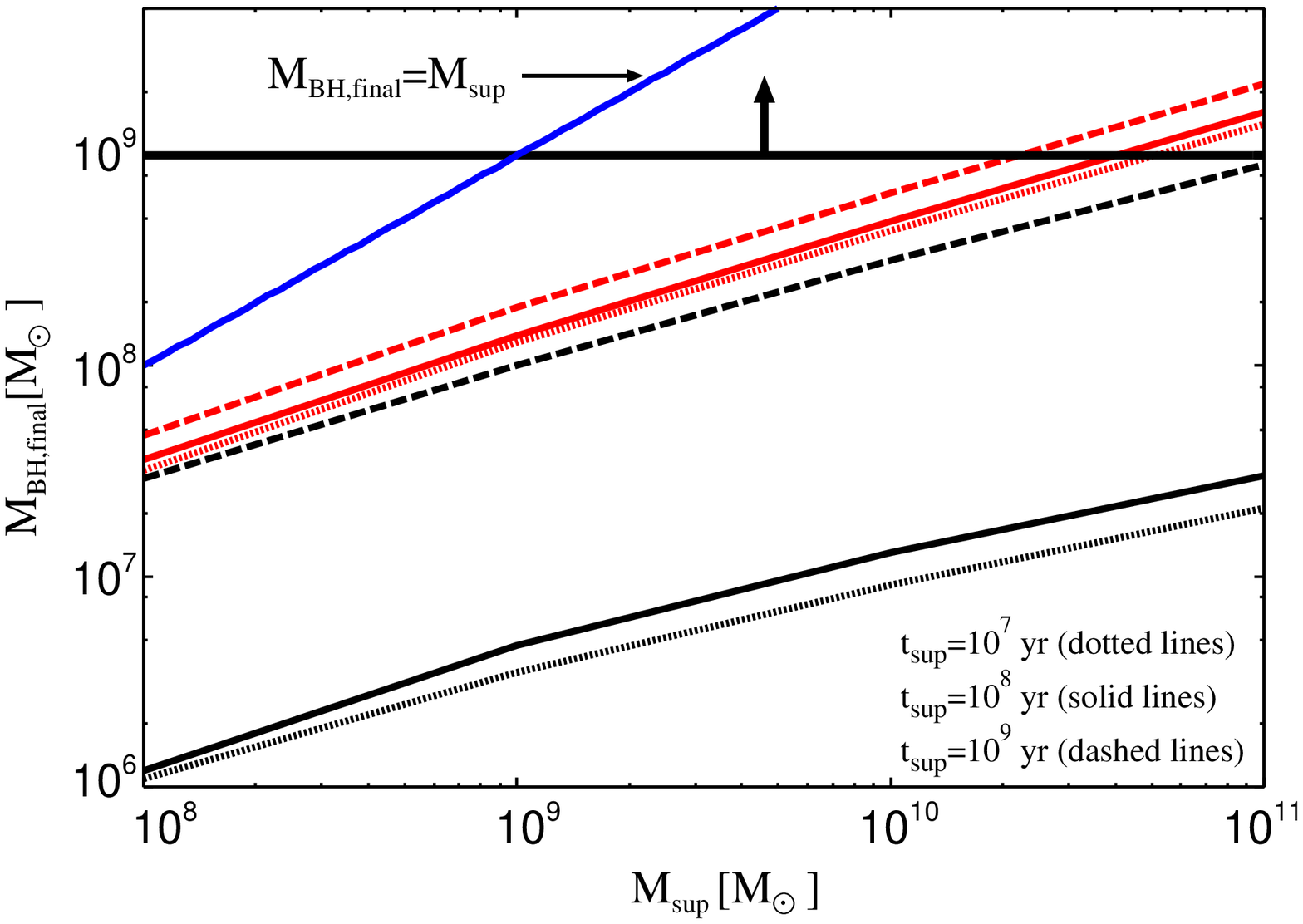}
\figcaption
{
The final SMBH mass, $M_{\rm BH, final}$ against 
the different total accreted mass from host galaxies, $M_{\rm sup}\equiv 
\dot{M}_{\rm sup}t_{\rm sup}$ for $t_{\rm sup}=10^{7}\,{\rm yr}$, 
$10^{8}\,{\rm yr}$ and $10^{9}\,{\rm yr}$. 
The red lines are the results of the super-Eddington growth model, 
while the black lines are ones of the Eddington-limitted growth model. 
The blue line denotes that all the supplied mass is used to grow the SMBH, 
i.e., $M_{\rm BH,final}=M_{\rm sup}$. 
Here we assume $C_{*}=3\times 10^{-8}\,{\rm yr}^{-1}$.
}
\vspace{2mm}

\section{Necessary conditions of formation of SMBHs at $z > 6$}
Based on the coevolution model of CNDs and SMBHs ($\S2$ and KW08), 
we examine conditions of the formation of SMBHs with $M_{\rm BH}
\geq10^{9}M_{\odot}$ at $z >6$. 
We first show results of the super-Eddington growth model in $\S 3.1$, and 
those of the Eddington-limited growth model in $\S 3.2$. 

\subsection{Super-Eddington growth model}
Figure 1 shows the final BH mass, $M_{\rm BH,final}$ as a function of 
the total mass accreted  from hosts, $M_{\rm sup}$ for the different period 
of mass supply from hosts, $t_{\rm sup}$ assuming $C_{*}=3\times 10^{-8}
\,{\rm yr}^{-1}$. This value is comparable to the star formation efficiency 
of high-$z$ galaxies (e.g., Tacconi et al. 2006; see also Fig. 5). 
The effect of $C_{*}$ will be discussed below. 
The red lines are the results of the super-Eddington growth model 
for $t_{\rm sup}=10^{7}\,{\rm yr}$, $10^{8}\,{\rm yr}$, and $10^{9}\,{\rm yr}$. The blue line denotes that all the supplied mass is used to grow the SMBH, 
i.e., $M_{\rm BH,final}=M_{\rm sup}$. 
It is found that $M_{\rm BH, final}\approx (0.01-0.1)M_{\rm sup}$. 
This low efficiency is independent of the period of the mass-supply 
from hosts, $t_{\rm sup}$ and the mass supply rate of hosts, 
$\dot{M}_{\rm sup}$. 
In order to form SMBHs with $> 10^{9}M_{\odot}$, plenty of infalling gas 
from host galaxies, i.e., $M_{\rm sup}\,> 10^{10}M_{\rm \odot}$ is requested. 
We consider why the mass ratio, $M_{\rm BH, final}/M_{\rm sup}$ is too small 
as we showed. 
From the BH growth rate for the high accretion rate 
($\S2.2$ and see also eq. (12) in KW08), the accretion energy onto 
a central BH, 
i.e., $0.5\dot{M}_{\rm BH}v_{\rm in}^{2}$ where $v_{\rm in}=\sqrt
{GM_{\rm BH}/r_{\rm in}}$, can be obtained as 
\begin{equation}
\frac{1}{2}\dot{M}_{\rm BH}v_{\rm in}^{2}=2f_{\rm \gamma}
\alpha_{\rm SN}\eta_{\rm SN} E_{\rm SN}\dot{M}_{*}, 
\end{equation}
where $f_{\gamma}\equiv [3(14-4\gamma)(8-4\gamma)/16]
(r_{\rm in}/r_{\rm out})^{2-\gamma}$. 
Thus, equation (7) indicates that $\alpha_{\rm SN}$ and $\eta_{\rm SN}$ 
are important parameters to determine the conversion efficiency 
from the energy input from SN explosions ($\approx E_{\rm SN}\dot{M}_{*}$) to the accretion 
energy onto a central BH ($\approx \dot{M}_{\rm BH}v^{2}_{\rm in}$). 
Using the fiducial values of $\gamma=1$, 
the ratio of the SMBH growth and star formation rate of CNDs is given by 
\begin{equation}
\frac{\dot{M}_{\rm BH}}{\dot{M}_{\rm *}}=
0.3\alpha_{\rm SN} \left(\frac{\eta_{\rm SN}}{10^{-3}M^{-1}_{\odot}}\right) \left(\frac{M_{\rm disk}}{10^{8}M_{\odot}}\right)^{-0.5}.
\end{equation}
This indicates that larger $M_{\rm sup}$ (or $M_{\rm disk}$) leads to smaller 
$M_{\rm BH, final}/M_{\rm sup}$. This can be understood as follows. 
As $M_{\rm sup}$ becomes larger $M_{\rm BH}$ increases. The larger 
BH mass causes a smaller scale height and a smaller turbulent velocity 
because the CND is vertically supported by the turbulent pressure (eq.(1)). 
Thus, the growth rate of BHs ($\propto \nu_{\rm t}\propto v_{\rm t}h$) 
decreases as $M_{\rm sup}$ (or $M_{\rm disk}$) is larger. 
Note that $M_{\rm disk}=M_{\rm g}+M_{*}$ is comparable to 
$M_{\rm sup}$. 
If we focus on the case for $M_{\rm sup} > 10^{8}M_{\odot}$, it turns out that 
the BH growth is always smaller than the star formation rate in the CND, 
i.e., $\dot{M}_{\rm BH} < 0.3\dot{M}_{*}$, assuming that $\alpha_{\rm SN}=1$ 
and $\eta_{\rm SN}=10^{-3}M_{\odot}^{-1}$
\footnote{
The numerical simulations have shown $\alpha_{\rm SN}\approx 1$ (e.g., WN02). 
On the other hand, the parameter $\eta_{\rm SN}$ is expressed as $\eta_{\rm SN}\equiv \epsilon_{\rm SN} f_{\rm SN}$, where $\epsilon_{\rm SN}$ and $f_{\rm SN}$ are the efficiency with which SN energy is transferred to the gas in the CND, and the number density of SNe per solar mass of the star formation, respectively. In this paper (also in KW08), we assume $\eta_{\rm SN}=10^{-3}M^{-1}_{\odot}$, 
that is, $\epsilon_{\rm SN}=0.1$ (e.g., Thornton et al. 1998; 
Wada \& Norman 2002; Wada et al. 2009), and $f_{\rm SN}=10^{-2}M^{-1}_{\odot}$ which is expected for Salpeter initial mass function (IMF) with low-mass cutoff being $m_{\rm l}=0.1M_{\odot}$. Note that the heating efficiency 
$\eta_{\rm SN}$ must be larger than $10^{-3}M_{\odot}^{-1}$ in order to be 
satisfied with $\dot{M}_{\rm BH} > \dot{M}_{*}$. 
}.
In other words, the star formation rate always overcomes the mass accretion rate onto BHs in this model. 
From the eq. (8), at given $M_{\rm sup}$ (or $M_{\rm disk}$) the ratio 
$\dot{M}_{\rm BH}/\dot{M}_{*}$ is basically determined by two quantities, 
i.e., $\alpha_{\rm SN}$ and $\eta_{\rm SN}$ which control turbulent viscosity 
and efficiency of SN heating, respectively. 

\vspace{5mm}
\epsfxsize=8cm 
\epsfbox{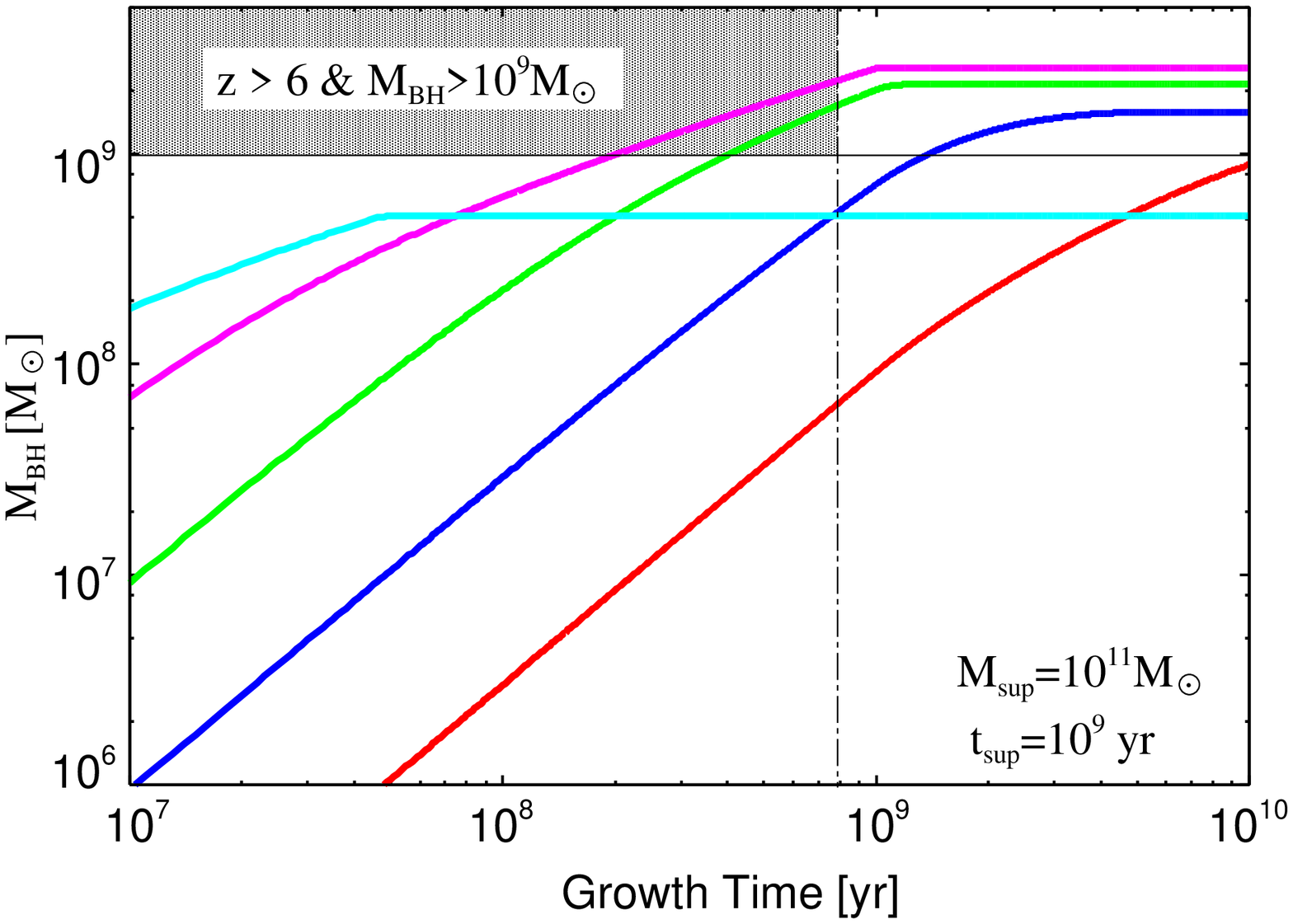}
\figcaption
{
Time evolution of the mass of BH for different $C_{*}$. 
The right blue, magenta, green, blue and red lines correspond to 
$C_{*}=3\times 10^{-6}\,{\rm yr}^{-1}$, $3\times 10^{-7}\,{\rm yr}^{-1}$, 
$3\times 10^{-8}\,{\rm yr}^{-1}$, $3\times 10^{-9}\,{\rm yr}^{-1}$ and 
$3\times 10^{-10}\,{\rm yr}^{-1}$, respectively. Here we assume $M_{\rm sup}
=10^{11}M_{\odot}$ and $t_{\rm sup}=10^{9}\,{\rm yr}$. 
The shaded region represents the allowed region for the formation of SMBHs 
with $M_{\rm BH}>10^{9}M_{\odot}$ at $z>6$. 
The dot-dashed vertical line corresponds to $z=6$ if the redshift $z_{\rm i}$ 
at which the accretion onto a seed BH starts is $z_{\rm i}=25$.
}
\vspace{2mm}
\vspace{5mm}
\epsfxsize=8cm 
\epsfbox{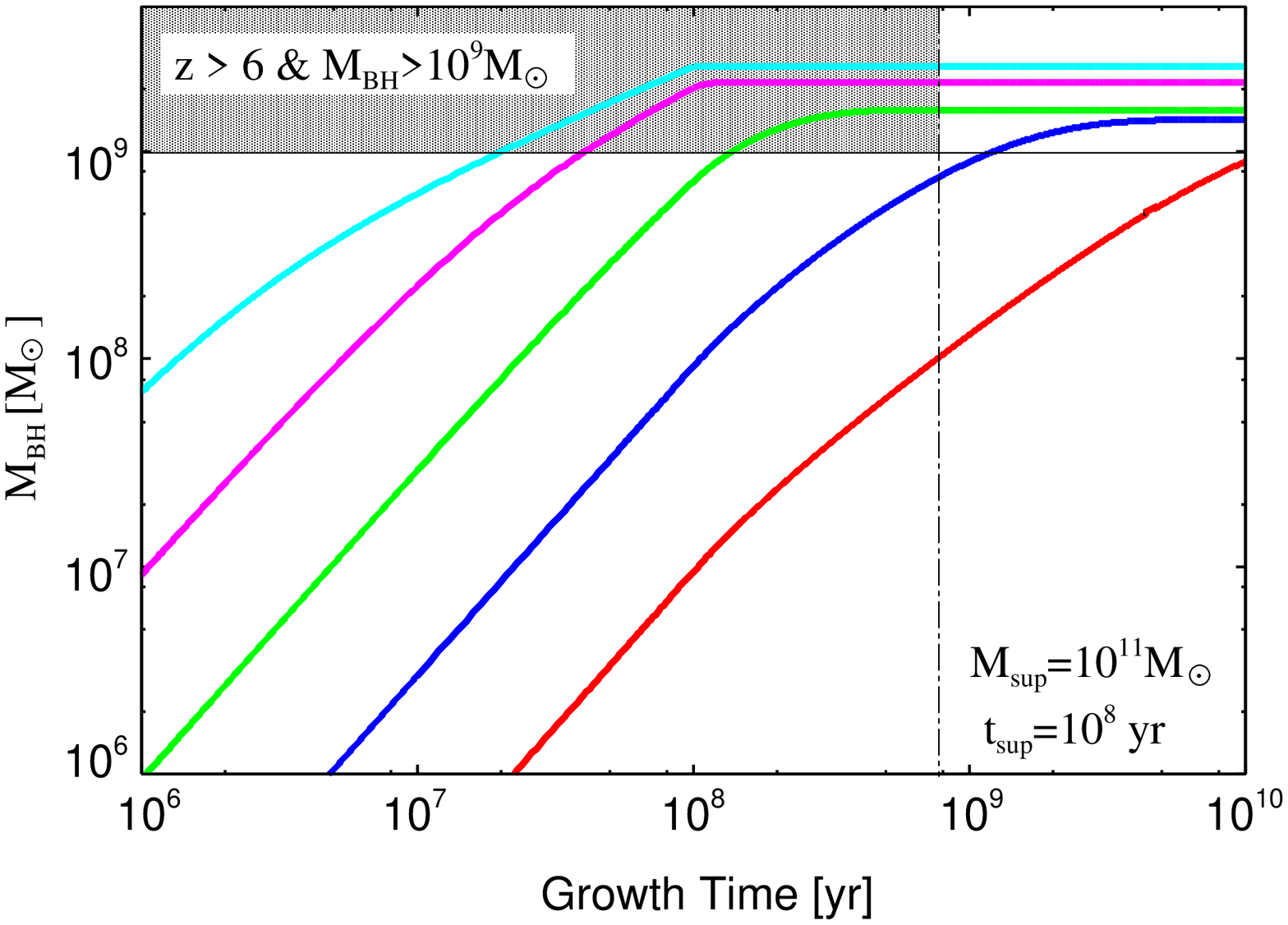}
\figcaption
{
Same as Fig. 2, but $t_{\rm sup}=10^{8}\,{\rm yr}$. 
The lines are same as Fig. 2.
}
\vspace{2mm}

Next, we examine the dependence of the star formation efficiency, $C_{*}
=\dot{M}_{*}/M_{\rm g}$ on the mass of SMBHs, $M_{\rm BH}(t)$. 
Figure 2 shows the evolution of the SMBH mass, $M_{\rm BH}(t)$ for 
the different star formation efficiency, $C_{*}$. 
We here assume the total mass accreted from hosts, $M_{\rm sup}=10^{11}M_{\odot}$ and a period of the mass-supply from hosts, $t_{\rm sup}=10^{9}\,{\rm yr}$. 
The redshift $z_{\rm i}$ when the accretion onto a seed BH got 
started is supposed to be $z_{\rm i}=25$. 
Thus, the growth timescale of SMBHs must be less 
than $8\times 10^{8}\,{\rm yr}$ (vertical dot-dashed line) to form the SMBH 
at $z >6$. 
The allowed region to form a SMBH with $M_{\rm BH} >10^{9}M_{\odot}$ 
at $z > 6$ is shown as shaded. 
We find that $M_{\rm BH}(t)$ significantly depends on the star formation 
efficiency, $C_{*}$. 
If the star formation efficiency is extremely low (e.g., $C_{*}\leq 
3\times 10^{-9}\,{\rm yr}^{-1}$), the timescale of SMBH 
growth is larger than the Hubble time at $z=0$ because of $t_{\rm growth}
\approx r^{2}_{\rm in}/v_{\rm t}h \sim (f_{\gamma}C_{*})^{-1}$ (see eq. (14) 
in KW08). Thus, $M_{\rm BH, final}$ cannot be larger than $\sim 
10^{9}M_{\odot}$ until $z=6$ (the red line in Fig. 2). 
For the extremely high efficiency (e.g., $C_{*}\geq 3\times 10^{-6}\,{\rm yr}^{-1}$; light blue line in Fig. 2), 
on the other hand, the gas supplied from the host is mostly consumed by 
the star formation ($\dot{M}_{\rm sup} < \dot{M}_{*}$), thus the final SMBH 
mass is smaller than $10^{9}M_{\odot}$. Therefore, in order to form a SMBH 
with $M_{\rm BH} > 10^{9}M_{\odot}$ at $z >6$, 
relatively high star formation efficiency, i.e., $10^{-8}\,{\rm yr}^{-1} \leq C_{*} \leq 10^{-7}\,{\rm yr}^{-1}$ is required.

We show the evolution of $M_{\rm BH}(t)$ for the case of $t_{\rm sup}=10^{8}\,
{\rm yr}$ (Fig. 3) and $t_{\rm sup}=5\times 10^{9}\,{\rm yr}$ (Fig. 4) 
respectively, to elucidate the dependence of a period of the mass-supply from 
hosts $t_{\rm sup}$ on $M_{\rm BH}(t)$. 
Comparing Fig. 2 with Fig. 3, the final BH masses 
for $t_{\rm sup}=10^{8}\,{\rm yr}$ are almost the same as that for 
$t_{\rm sup}=10^{9}\,{\rm yr}$ (Fig. 2), but the behavior of $M_{\rm BH}(t)$ 
for $t_{\rm sup}=10^{8}\,{\rm yr}$ is different from that for $t_{\rm sup}
=10^{9}\,{\rm yr}$ at given $C_{*}=3\times 10^{-6}\,{\rm yr}^{-1}$ 
(the light blue line). 
This is because for $t_{\rm sup}=10^{8}\,{\rm yr}$ the mass supplied rate 
is higher than the star formation rate ($\dot{M}_{\rm sup} > \dot{M}_{*}$). 
Thus, the final BH mass can be as large as $\sim 10^{9}M_{\odot}$. 
As a result, the allowed region is $10^{-8}\,{\rm yr}^{-1} \leq C_{*} \leq 10^{-6}\,{\rm yr}^{-1}$ for $t_{\rm sup}=10^{8}\,{\rm yr}$. 
The upper value $C_{*}$ is determined by the local gravitational free-fall time scale, $t_{\rm ff}\approx (Gn_{\rm disk}m_{\rm p})^{-1/2}\sim 10^{6}$ yr 
for the averaged density of CNDs, i.e., $n_{\rm disk}=10^{3-4}\,{\rm cm}^{-3}$, because of the star formation time scale $t_{*}=C_{*}^{-1}\geq t_{\rm ff}$. 
We also found that the required range of $C_{*}$ for $t_{\rm sup}=
10^{7}\,{\rm yr}$ is the same as those for $t_{\rm sup}=10^{8}\,{\rm yr}$.
Figure 4 shows that there is no solution to form the SMBH with $M_{\rm BH} > 10^{9}M_{\odot}$ at $z >6$ for $t_{\rm sup}=5\times 10^{9}\,{\rm yr}$ 
because of $\dot{M}_{\rm sup} < \dot{M}_{*}$ for the wide range of $C_{*}$, 
i.e., $3\times 10^{-10}\,{\rm yr}^{-1}\leq C_{*} \leq 3\times 10^{-6}\,{\rm yr}^{-1}$. In other words, this indicates the rapid mass supply from the host to the central tens pc region ($\dot{M}_{\rm sup} > 20M_{\odot}\,{\rm yr}^{-1}$, 
$M_{\sup}=10^{11}M_{\odot}$) is also necessary to  form the SMBHs at $z >6$.

\vspace{2mm}
\vspace{5mm}
\epsfxsize=8cm 
\epsfbox{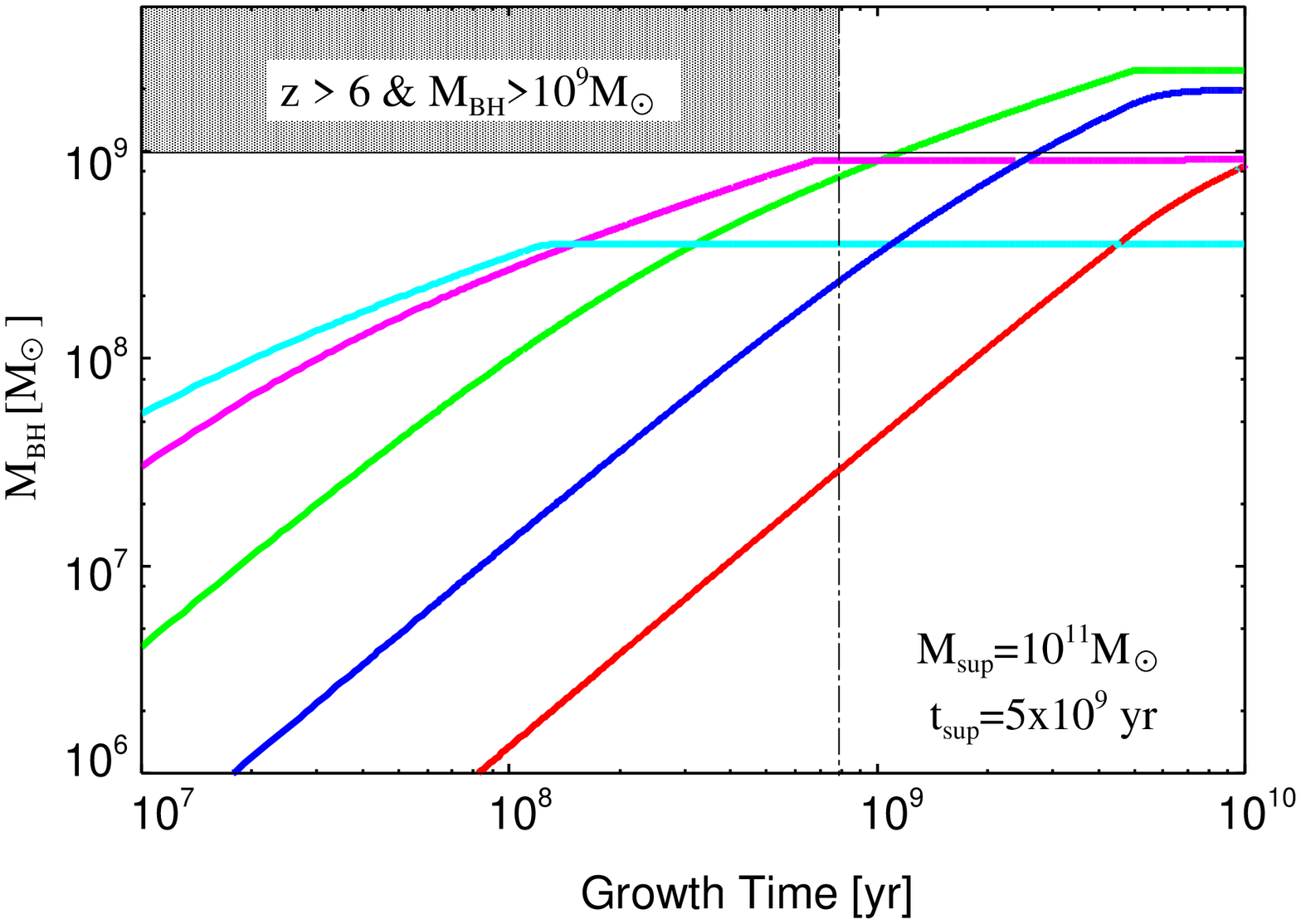}
\figcaption
{
Same as Fig. 2, but for $t_{\rm sup}=5\times 10^{9}\,{\rm yr}$. 
The lines are same as Fig. 2. 
}
\vspace{2mm}

Figure 5 shows the SFR as a function of the surface density of gas in CNDs, 
$\Sigma_{\rm g}$. 
We estimate the surface gas density of CNDs as $\Sigma_{\rm g}=M_{\rm g}/
\pi r^{2}_{\rm out}$. Here we take $M_{\rm g}$ and $r_{\rm out}$ from 
our calculations. The size of the CND increases with time but finally 
$r_{\rm out}$ is $\approx 2\times 10^{3}$ pc. 
For the SFR and $\Sigma_{\rm g}$ in the shaded region, SMBHs with 
$M_{\rm BH,final}>10^{9}M_{\odot}$ at $z >6$ can be formed as a result of 
considering the dependences of $C_{*}$ and $\Sigma_{\rm g}$. 
The observational data of nearby normal galaxies 
(Komugi et al. 2005) and nearby starburst galaxies (Kennicutt 1998) 
are plotted in Fig. 5. 
The two plots represent the average of massive star forming submillimeter 
galaxies (SMGs) at $z\approx 2.5$ with ${\rm SFR}\approx 10^{-4}M_{\odot}\,{\rm yr}^{-1}\,{\rm pc}^{-2}$ and $\Sigma_{\rm g}\approx 5\times 10^{3}M_{\odot}\,{\rm pc}^{-2}$ (e.g., Tacconi et al. 2006) and a QSO at $z=6.42$ (SDSS J1148) with ${\rm SFR}\approx 10^{-3}M_{\odot}\,{\rm yr}^{-1}\, {\rm pc}^{-2}$ and $\Sigma_{\rm g}\approx 10^{4}M_{\odot}\,{\rm pc}^{-2}$  (Walter et al. 2009). 
The required $C_{*}$ for high-$z$ SMBH formation is comparable to 
SMGs and QSO hosts at $z=6.42$, although this is higher than that of 
low-$z$ normal galaxies and starburst ones. 
Thus, the predicted high star formation efficiency, i.e., 
$10^{-8}\,{\rm yr}^{-1}\leq C_{*} \leq 10^{-6}\,{\rm yr}^{-1}$ 
actually occurred in the high-$z$ universe. 

\vspace{5mm}
\epsfxsize=8cm 
\epsfbox{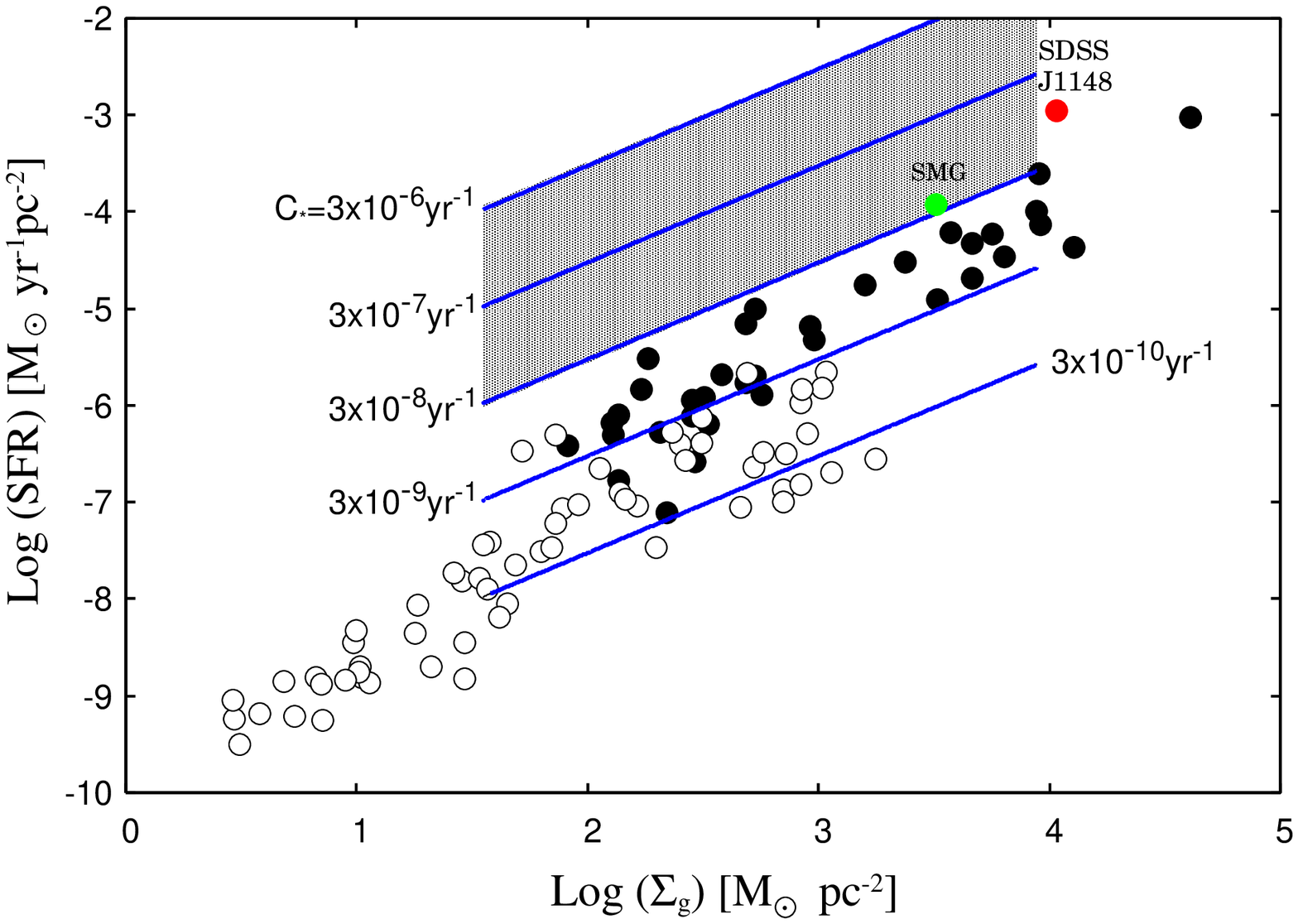}
\figcaption
{
Comparison between the observed surface star formation rate (SFR) in terms of 
the surface gas density ($\Sigma_{\rm g}$) and the allowed region for the formation of SMBHs with $M_{\rm BH}>10^{9}M_{\odot}$ at $z >6$ (the shaded region). 
The blue lines represent theoretical predictions for different $C_{*}$, 
assuming $M_{\rm sup}=10^{11}M_{\odot}$ and $t_{\rm sup}=10^{8}\,{\rm yr}$. 
Black filled dots are starburst galaxies in Kennicutt (1998) and the 
black open dots are normal galaxies in Komugi et al. (2005). 
Green point denotes the average dusty submillimeter galaxies (SMGs) at 
$z\approx 2.5$ (Tacconi et al. 2006) and red point represents $\sim$ kpc scale 
starforming region around the $z=6.42$ SDSS J1148+5251 (Walter et al. 2009).
}
\vspace{2mm}

In summary, in order to form the high-$z$ SMBH 
(i.e., $z >6$ and $M_{\rm BH,final} > 10^{9}M_{\odot}$) 
by super-Eddington gas accretion  
we need;
(i) abundant mass supply from the galactic scale ($\sim$ kpc), i.e., 
$M_{\rm sup} > 10^{10}M_{\odot}$, 
(ii) rapid mass inflow from the galactic scale ($\sim$ kpc), i.e., 
$t_{\rm sup} < 5\times 10^{9}\,{\rm yr}$ and 
(iii) high star formation efficiency, i.e., 
$10^{-8}\,{\rm yr}^{-1}\leq C_{*} \leq 10^{-6}\,{\rm yr}^{-1}$.

\subsection{Eddington-limited growth model} 
We here discuss the results of Eddington-limited growth model 
(the black lines in Fig. 1). 
Figure 1 shows that the Eddington-limited growth model leads to 
much lower efficiency for $t_{\rm sup}=10^{7}\,{\rm yr}$ and $10^{8}\,
{\rm yr}$, i.e., $M_{\rm BH, final}\approx (10^{-4}-10^{-2})M_{\rm sup}$. 
But, the efficiency for $t_{\rm sup}=10^{9}\,{\rm yr}$ is comparable to 
that of the super-Eddington growth model, i.e, $M_{\rm BH, final}\approx 
(0.01-0.1)M_{\rm sup}$. These results can be understood in terms of 
the ratio of the mass-supply time scale, $t_{\rm sup}$ and 
the Eddington time scale, $t_{\rm Edd}=\epsilon_{\rm BH}\sigma_{\rm T}c
/(4\pi Gm_{\rm p})$. 
The black hole grows exponentially at the late time ($t\gg t_{\rm Edd}
\approx 5\times 10^{7}\,{\rm yr}$ for $\epsilon_{\rm BH}=0.1$) 
because of $\dot{M}_{\rm BH}\propto e^{t/t_{\rm Edd}}$. 
If $t_{\rm sup}$ is comparable and/or shorter than $t_{\rm Edd}$ 
(e.g., $t_{\rm sup}=10^{7}\,{\rm yr}$ and $10^{8}\,{\rm yr}$), 
the mass-supply from a host galaxy terminates before a main growth phase of BHs. Thus, the BH growth is greatly suppressed by the AGN outflow (see eq. (6)), 
and then the final BH mass can be 1-2 orders of magnitude smaller than 
that of the super-Eddington growth model as seen in Fig. 1. 
On the other hand, the effect of AGN outflows in high accretion phase 
is not significant for $t_{\rm sup}=10^{9}\,{\rm yr}$ 
because of $t_{\rm sup}\gg t_{\rm Edd}$. Thus, the final BH mass is 
comparable to that of the super-Eddington growth model. 
In order to form a SMBH with $> 10^{9}M_{\odot}$, it is found that 
the super-Eddington growth is needed when $t_{\rm sup}$ is 
shorter than $\sim 10^{9}\,{\rm yr}$. 
Note that the dependence of $C_{*}$ and $t_{\rm sup}$ on the 
final BH mass are the same as those for the super-Eddington growth model.

\section{Observable properties and evolution of QSOs at $z >6$}
Based on the present model in $\S 3$, we here discuss observable properties and the evolution of QSOs at $z >6$, in order to compare them with the current and 
future observations. 
It is still unclear when the accretion onto a seed BH got started in the high
-$z$ universe. 
Recently, Johnson \& Bromn (2007) mentioned that the radiation feedback 
from the first stars may deplete the gas in the central region of the 
proto-galaxy and may delay the BH accretion by up to 
$\sim 10^{8}\,{\rm yr}$. 
Thus, we explore the relationship between the evolution of QSOs at $z > 6$ 
and the redshift $z_{\rm i}$ when the seed BH begins to grow. 
To this end, we examine the evolution and formation of high-$z$ QSOs 
at $z >6$ for the two scenarios of high-$z$ QSO formation; 
(a) {\it the early growth of SMBHs}, i.e., $z_{\rm i}=25$. 
This corresponds to the supply period $t_{\rm sup}=7\times10^{8}\,{\rm yr}$ 
and (b) {\it the late growth of SMBHs}, i.e., $z_{\rm i}=8$. 
This corresponds to the supply period $t_{\rm sup}=10^{8}\,{\rm yr}$. 
We here use the fiducial values, i.e., $M_{\rm sup}=10^{11}M_{\odot}$ and 
$C_{*}=3\times 10^{-8}\,{\rm yr}^{-1}$, for which a SMBH with $M_{\rm BH} 
>10^{9}M_{\odot}$ at $z >6$ can be formed. 
We define the QSO phase as $t_{\rm sup} < t < t_{\rm QSO}$ where $t_{\rm QSO}$ 
is the time when the AGN luminosity equals to the threshold luminosity of QSOs, $L_{\rm QSO, th}$. We assume $L_{\rm QSO, th}=10^{46}\,{\rm erg}\,{\rm s}^{-1}$. The proto-QSO phase is defined as the early phase of a growing BH 
($0 \leq t \leq t_{\rm sup}$). 

\vspace{5mm}
\epsfxsize=8cm 
\epsfbox{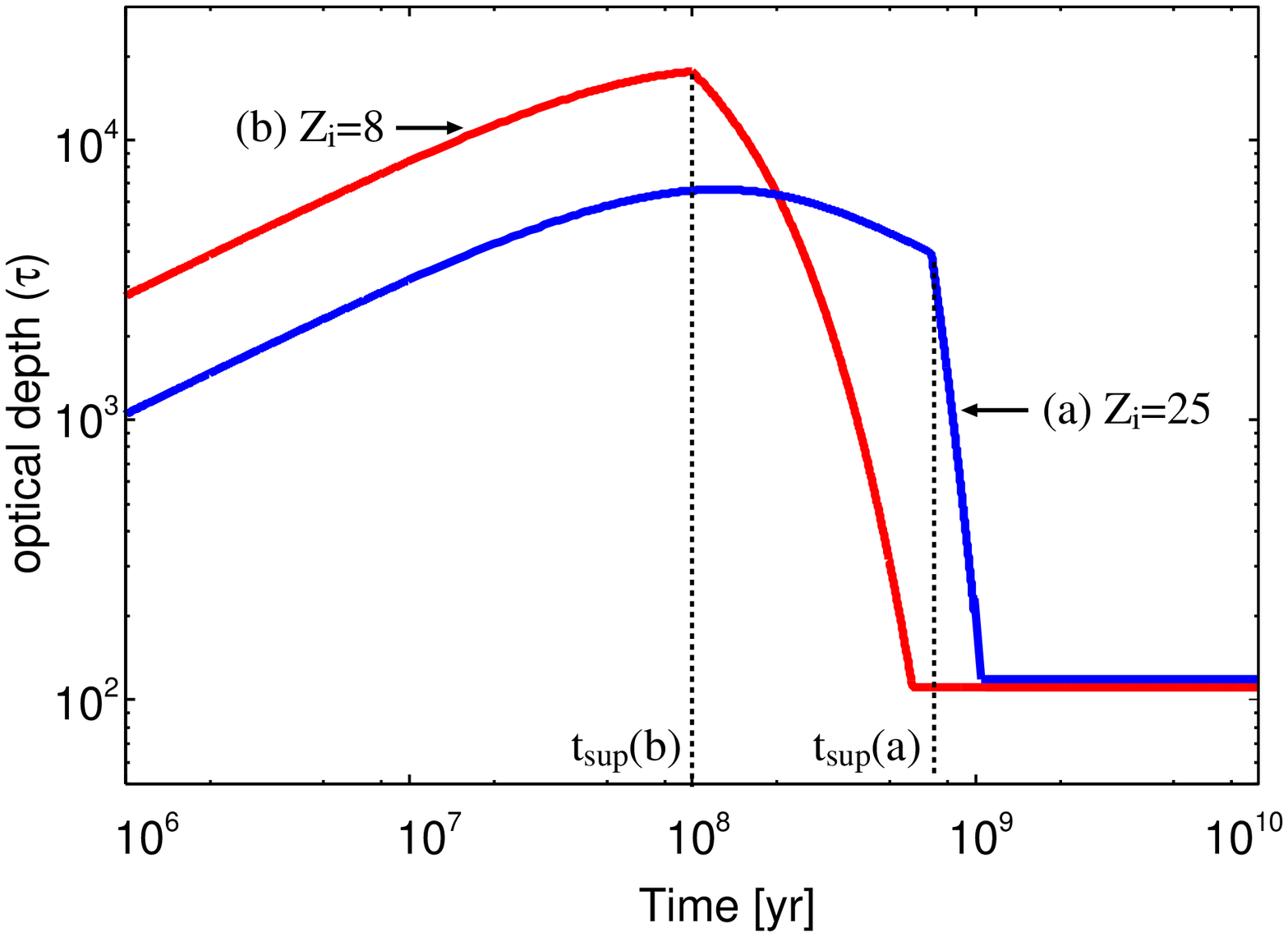}
\figcaption
{
Time evolution of optical depth of CNDs for UV-band in the edge-on view 
($\tau$). 
The blue line corresponds to the case for $z_{\rm i}=25$ (or $t_{\rm sup}
{\rm (a)}=7\times 10^{8}\,{\rm yr}$). 
The red line denotes the case for $z_{\rm i}=8$ (or $t_{\rm sup}{\rm (b)}=10^{8}\,{\rm yr}$).
}
\vspace{2mm}

Figure 6 shows the evolution of optical depth of CNDs for UV-band 
in the edge-on view, 
$\tau (t)$ 
defined as $\tau (t)\equiv \int^{r_{\rm out}}_{r_{\rm in}}\chi_{\rm d}
\rho_{\rm g}(r)dr\simeq \chi_{\rm d} M_{\rm g}(r)/[2\pi r_{\rm out}^{2}(r)
h_{\rm out}(r)]$ where $\chi_{\rm d}$ is the mass extinction due to the dust 
and $h_{\rm out}(r)$ is the scale height at $r_{\rm out}(r)$. 
The mass extinction coefficient $\chi_{\rm d}$ is given by 
$\chi_{\rm d}=n_{\rm d}\sigma_{\rm d}/
\rho_{\rm g}$ with the number density $n_{\rm d}$ and cross section 
$\sigma_{\rm d}$. 
In this paper, we assume $\chi_{\rm d}=10^{3}\,{\rm cm}^{2}\, {\rm g}^{-1}
(a_{\rm d}/0.1\,\mu{\rm m})^{-1}(\rho_{\rm s}/{\rm g}\,{\rm cm}^{-3})(Z/Z_{\odot})$, 
where $a_{\rm d}$ is the grain radius, $\rho_{\rm s}$ is the density of 
solid material density within the grain (e.g., Spitzer 1978; their $\S$9.3), 
and $Z$ is the metallicity of gas, which are fixed at $a_{\rm d} = 0.1\mu m$, 
$\rho_{\rm s} =1 {\rm g}\,{\rm cm}^{-3}$, and $Z=Z_{\odot}$. 
We find that CNDs are always optically thick during the evolutionary path of 
bright QSOs for both cases (i.e., $z_{\rm i}=8$ and $25$). 
The present model predicts the existence of an optically thick CND even in a 
proto-QSO phase. 
At late times ($t > 10^{9}$ yr), the properties of CNDs (e.g., $M_{\rm g}$, $r_{\rm out}$, and $h_{\rm out}$) are independent of time because of 
no fueling and no star formation in this phase. Thus, the optical depth is 
constant in time.

In order to investigate the evolution of the AGN luminosity and the 
nuclear-starburst luminosity, we define these luminosities as follows:
Following Watarai et al. (2000), the AGN luminosity can be given as a function 
of $\dot{m}_{\rm BH}\equiv \dot{M}_{\rm BH}/\dot{M}_{\rm Edd}$ 
\begin{equation}
L_{\rm AGN}(t)=\left \{
 \begin{array}{l}
 2\left(1+\ln{\frac{\dot{m}_{\rm BH}(t)}{20}}\right)
L_{\rm Edd}(t)\,\,\, ;\dot{m}_{\rm BH}(t) \geq 20, \\ \\
 \left(\frac{\dot{m}_{\rm BH}(t)}{10}\right)
L_{\rm Edd}(t) \,\,\,\,\, ;\dot{m}_{\rm BH}(t) < 20, 
 \end{array}\right .
\end{equation}
where $\dot{m}_{\rm BH}\equiv \dot{M}_{\rm BH}/\dot{M}_{\rm Edd}$. 
On the other hand, the nuclear starburst luminosity $L_{\rm SB}(t)$ 
can be obtained as
\begin{equation}
L_{\rm SB}(t)=0.14\epsilon_{*} \dot{M}_{*}(t)c^{2}, 
\end{equation}
where $\epsilon_{*}=0.007$ which is the energy conversion efficiency of 
nuclear fusion from hydrogen to helium. 
The results of the super-Eddington growth model are shown in $\S 4.1$. 
Results based on the Eddington-limited accretion are discussed on 
$\S 4.2$. 

\subsection{Super-Eddington growth model}
Figure 7 (a) and (b) shows the evolution of bolometric luminosity of AGN, 
$L_{\rm AGN}$ (red lines), nuclear starburst luminosity, $L_{\rm SB}$ 
(green lines) and Eddington luminosity, $L_{\rm Edd}$ (blue lines) 
for two scenarios of high-$z$ QSO formation. The shaded region shows QSO phase. 
We find that the evolution of QSO luminosity depends on the redshift 
$z_{\rm i}$ at which accretion onto a seed BH is initiated, 
although the final BH mass, $M_{\rm BH, final}$ and 
the evolution of BH mass, $M_{\rm BH}$ are independent of $z_{\rm i}$ 
(or $t_{\rm sup}$) at given $C_{*}$ (see Fig.1, 2 and 3). 
The time when $L_{\rm AGN}$ becomes maximum is at $t\simeq t_{\rm sup}$ 
for $z_{\rm i}=8$, while it reaches the maximum much earlier than the QSO phase, i.e., $t_{\rm max}\sim 10^{8} \,{\rm yr}<t_{\rm sup}$, for $z_{\rm i}=25$. 
In addition, the maximum luminosity of AGNs for $z_{\rm i}=8$ is 
an order of magnitude larger than that for $z_{\rm i}=25$. 
These results indicate that the evolution of the QSO luminosity is sensitive 
to the redshift $z_{\rm i}$ when the accretion onto a seed BH is initiated. 
In other words, it is easier to build-up bright QSOs if the gas is supplied 
rather late on a relatively short timescale. 
This can be understood as follows. 
For $z_{\rm i}=8$, the mass supply time scale, $t_{\rm sup}$ is shorter than 
the star formation time scale, $t_{*} \propto C_{*}^{-1}$. 
Thus, the gas in the CNDs can accrete onto a BH more efficiently, 
compared with the case for $z_{\rm i}=25$. 
On the contrary, if the average gas supply rate is smaller in  a longer period,
 (i.e. $z_{\rm i} =25$), the supplied gas is mainly consumed to form stars not to form a SMBH because of $t_{\rm sup} > t_{*}$. 

\vspace{5mm}
\epsfxsize=8cm 
\epsfbox{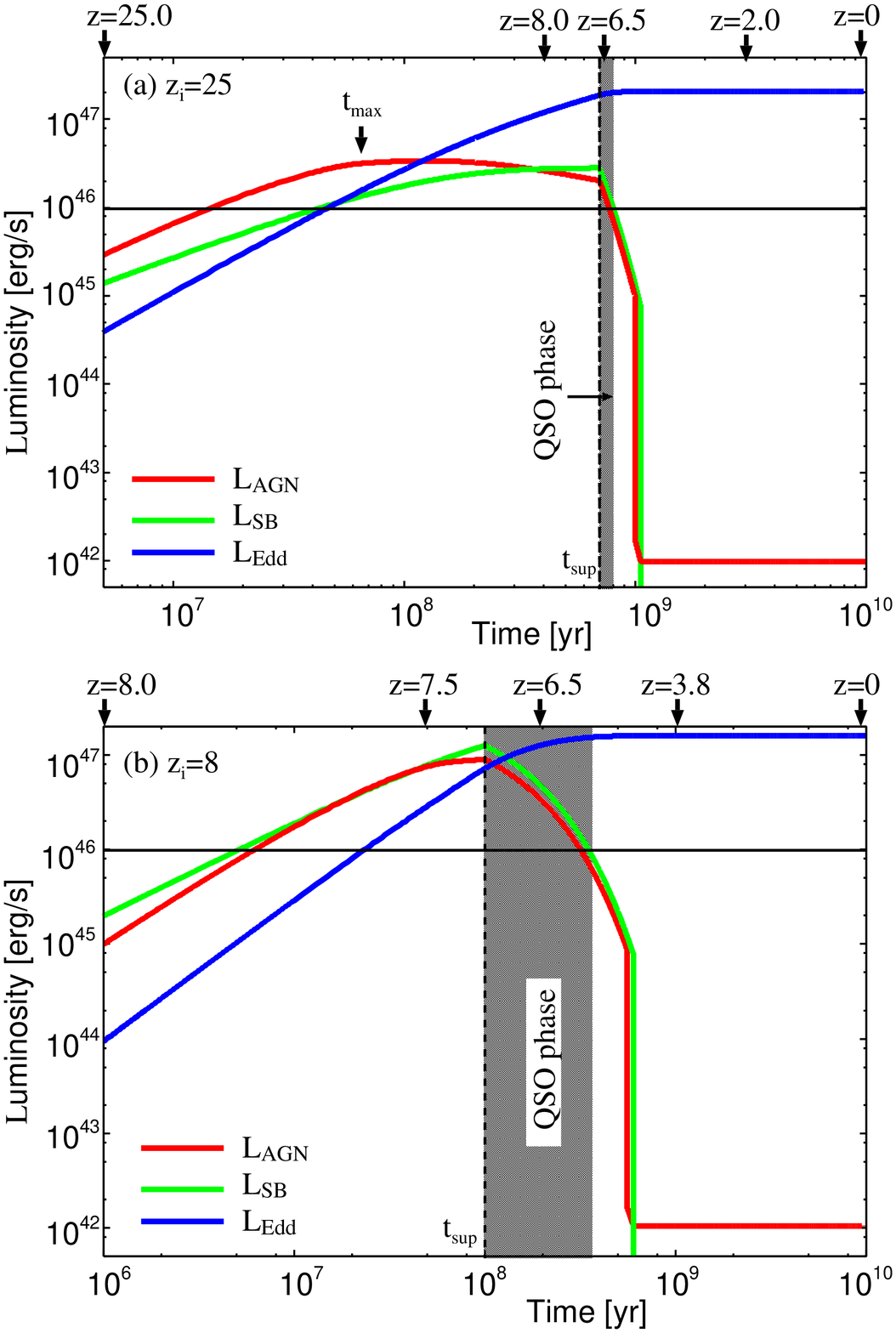}
\figcaption
{
(a) Time evolution of the AGN luminosity, $L_{\rm AGN}(t)$, that of 
the nuclear starburst luminosity, $L_{\rm SB}(t)$ and that of 
Eddington luminosity, $L_{\rm Edd}(t)$, assuming $M_{\rm sup}=10^{11}M_{\odot}$, $C_{*}=3\times 10^{-8}\,{\rm yr}^{-1}$ and $z_{\rm i}=25$ (or $t_{\rm sup}=7
\times 10^{8}\,{\rm yr}$). 
We define QSO phase as $t_{\rm sup} < t < t_{\rm QSO}$ (the shaded region), 
where $t_{\rm QSO}$ is the time when the AGN luminosity equals the 
threshold luminosity of AGN, $L_{\rm QSO,th}=10^{46}\,{\rm erg/s}$. 
For $M_{\rm BH}$ growth, we assume the super-Eddington growth model. 
(b) Same as (a), but for $z_{\rm i}=8$ (or $t_{\rm sup}=10^{8}\,{\rm yr}$). 
}
\vspace{2mm}

We also find that the AGN luminosity is sub-Eddington with 
$L_{\rm AGN}/L_{\rm Edd}\approx 0.1-1$ in the QSO phase, while 
it is super-Eddington with $L_{\rm AGN}/L_{\rm Edd}\approx 1-10$ 
in the proto-QSO phase (see Fig. 7(a) and (b)). 
On the other hand, the luminosity for the QSO phase 
is comparable to than that in the proto-QSO phase. 
This indicates that the Eddington ratio, $L_{\rm AGN}/L_{\rm Edd}$, 
varies two orders of magnitudes for a given AGN luminosity. 
This implies that the BH mass derived from only AGN luminosity 
leads to a large uncertainty.

The nuclear starburst (SB) luminosity for the proto-QSO phase is 
larger than that for the latter half of QSO phase, because the SB luminosity 
increases with time in the proto-QSO, while it decreases with time in the 
QSO phase. 
This trend is prominent for $z_{\rm i}=8$ rather than for $z_{\rm i}=25$ 
(see Fig.7 (a) and (b)). 
In this model, the enhanced star formation in a BH growing phase 
(i.e., a proto-QSO phase) is expected. 
At low-$z$ universe, Sani et al. (2009) discovered that the 
star formation activity in narrow line Seyfert galaxies (NLS1s) 
is larger than that in broad line Seyfert galaxies (BLS1s) of 
the same AGN luminosity. If NLS1s are the early phase of BLSIs, 
these observations are consistent with our predictions. 
We also find that the nuclear SB luminosity is comparable to 
the AGN luminosity in both the proto-QSO and QSO phases. 
Since young stars are buried in the optically thick CNDs (see Fig. 6), 
the intense nuclear starburst with $L_{\rm SB}\approx 10^{46}-10^{47}
\,{\rm erg/s}$ can be observed in the infrared band.

The lifetime of QSO is $\tau_{\rm QSO}\approx 1\times 10^{8}\, {\rm yr}$ 
for $z_{\rm i}=25$ and $\tau_{\rm QSO}\approx 3\times 10^{8}\, {\rm yr}$ 
for $z_{\rm i}=8$. 
These values of $\tau_{\rm QSO}$ are consistent with the previous studies 
(e.g., Martini et al. 2003a, b; Kawakatu, Umemura \& Mori 2003; Granato et al. 
2004; Shen et al. 2007; Li et al. 2008). 
From comparison between two models, we find that the QSO phase is longer 
in more luminous QSOs. 
We should note that $\tau_{\rm QSO}$ is larger for less luminous QSOs, 
e.g., $\tau_{\rm QSO}\approx 2\times 10^{8}\, {\rm yr}$ (case (a)) and 
$\tau_{\rm QSO}\approx 5\times 10^{8}\, {\rm yr}$ (case (b)) for 
$L_{\rm QSO, th}=10^{45}\,{\rm erg}\,{\rm s}^{-1}$. 
This has already been pointed out by Hopkins et al. (2005) and 
Li et al. (2007).

Figure 8 shows the time evolution of the BH mass, $M_{\rm BH}(t)$, 
the gas mass in the CND, $M_{\rm g}(t)$, the stellar mass in the disk, 
$M_{*}(t)$ and the total supplied mass from hosts. 
The evolution of $M_{\rm g}(t)$ is closely related to that of 
$L_{\rm AGN}(t)$ because the CNDs become gravitationally stable 
as $M_{\rm g}$ decreases (see $\S 2$). 
In the QSO phase, we find that $M_{\rm g}\approx (3-5)M_{\rm BH}=
5\times 10^{9}-1\times 10^{10}M_{\odot}$, $M_{*}\approx 100
M_{\rm BH}=10^{11}M_{\odot}$ for $z_{\rm i}=25$ (Fig. 8 (a)), and 
$M_{\rm g}\approx (3-30)M_{\rm BH}=5\times 10^{9}-
5\times 10^{10}M_{\odot}$ and $M_{*}\approx (10-100)M_{\rm BH}
=10^{10}-10^{11}M_{\odot}$ for $z_{\rm i}=8$ (Fig. 8 (b)).
Thus, it is predicted that there exists a {\it stellar rich 
massive CND} around SMBHs, i.e., $M_{\rm g}\approx 0.1M_{\rm dyn}$ 
where $M_{\rm dyn}$ is the dynamical mass of the CND 
plus BH system. 
On the other hand, in the proto-QSO phase it is found that 
$M_{\rm g}/M_{\rm BH}=10-10^{3}$ for $z_{\rm i}=25$ and $M_{\rm g}/M_{\rm BH}
=10^{2}-10^{3}$ for $z_{\rm i}=8$. 
Thus, we predict that the proto-QSOs have a {\it gas rich} CND around 
SMBHs, i.e., $M_{\rm g}\simeq M_{\rm dyn}$. 
At late times ($t > 10^{9}\,{\rm yr}$),  a very massive stellar nuclear 
disk ($M_{*}\sim 10^{11}M_{\odot}$) remains as a by-product of 
the formation of SMBHs with $M_{\rm BH}>10^{9}M_{\odot}$. 
Since the size of torus ($r_{\rm out}$) grows up $\sim 2$ kpc in this 
phase, the circular velocity of stellar components in the CND is $\sim 300\, 
{\rm km}\,{\rm s}^{-1}$. This is comparable to the velocity dispersion of 
giant elliptical galaxies (e.g., Tremaine et al. 2002). 
Comparing the case of $z_{\rm i}=25$ with that of $z_{\rm i}=8$, 
the larger amount of gas in the CND for $z_{\rm i}=8$ exists 
just before the QSO phase (see Fig. 8 (a) and (b)). 
As a result, the higher mass accretion rate ($\dot{M}_{\rm BH}$) is 
achieved, and the brighter AGN luminosity is sustained for a longer 
timescale. This is why the QSO lifetime for $z_{\rm i}=8$ 
is longer (see Fig. 7 (a) and (b)) than that for $z_{\rm i}=25$. 
In addition, for $z_{\rm i}=25$ the stellar mass $M_{*}$ becomes larger 
than the gas mass $M_{\rm g}$ at $t_{\rm max}\approx 10^{8}\,{\rm yr}$. 
Thus, the AGN luminosity does not increase after $t_{\rm max}
\approx 10^{8}\,{\rm yr}$ (see Fig. 7 (a)), even when the gas continues to 
infall from hosts into CNDs. This supplied gas is mainly consumed to form 
stars, not to form SMBH.

Based on the present results, we can predict the observable properties 
of proto-QSOs as follows; 
(i) a {\it gas rich} CND with the mass ratio $M_{\rm g}/M_{\rm BH}\geq 10^{2-3}$, which is larger than that of QSOs, i.e., $M_{\rm g}/M_{\rm BH}\leq 10$. 
(ii) the nuclear SB luminosity is slightly higher than that of 
QSOs. Thus, in order to explore proto-QSOs, it is essential to estimate 
the gas and stellar mass in CNDs, and the nuclear SB activity. 
Jiang et al. (2008, 2009) have recently discovered two magnitudes fainter 
QSOs at $z\sim 6$ than bright QSOs discovered in previous survey (e.g., Fan et al. 2003). Moreover, Kurk et al. (2009) found that 
two faint QSOs have very narrow broad emission lines (Mg II and CIV), 
which may imply small BH mass and high Eddington luminosity ratios. 
Thus, these faint QSOs may be good candidate of proto-QSOs with a 
super-Eddington mass accretion flow. 
It is interesting to examine the presence of gas rich CNDs for these 
high-$z$ faint QSOs using ALMA. 
But, we should mention that it is hard to judge whether the proto-QSOs are 
super-Eddington objects by only these observable properties of CNDs. 
Thus, we will really need to explore the direct evidence of 
super-Eddington objects. 

\vspace{5mm}
\epsfxsize=8cm 
\epsfbox{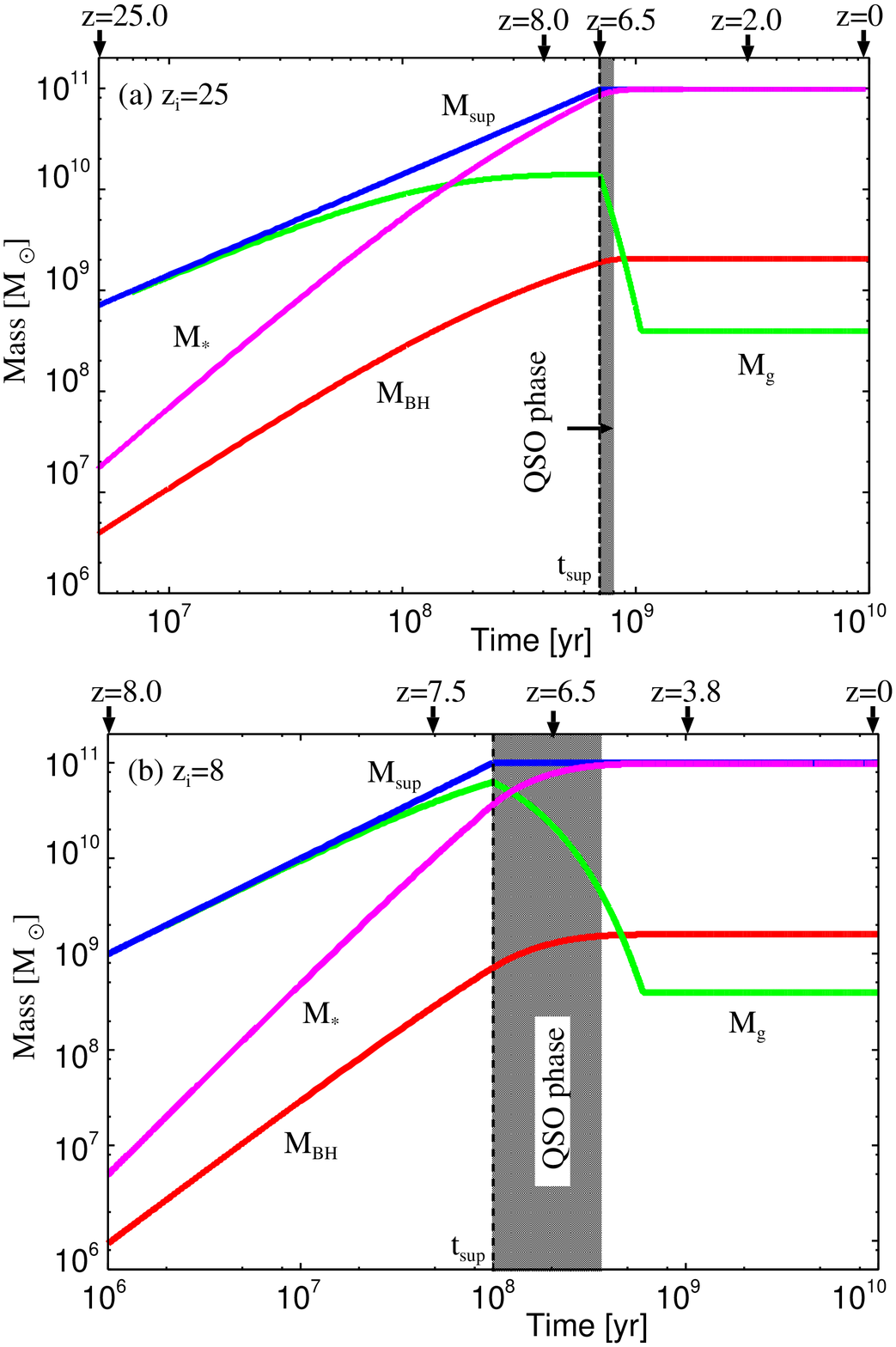}
\figcaption
{
(a) Time evolution of the BH mass, $M_{\rm BH}(t)$, the gas mass in the 
CND, $M_{\rm g}(t)$, the stellar mass in the disk, $M_{*}(t)$ and the total 
mass supplied  from hosts, $M_{\rm sup}(t)$ for $z_{\rm i}=25$ (or 
$t_{\rm sup}=7\times 10^{8}\,{\rm yr}$). 
The shaded region denotes the QSO phase. 
For $M_{\rm BH}$ growth, we assume the super-Eddington growth model. 
(b) Same as (a), but for $z_{\rm i}=8$ (or $t_{\rm sup}=10^{8}\,{\rm yr}$).
}
\vspace{2mm}

\subsection{Eddington-limited growth model}
In $\S 4.1$, we have shown the evolution of luminosity and mass 
for the super-Eddington growth model (see Fig. 7 and Fig. 8). 
In order to compare with the super-Eddington growth model, 
we examine these evolutions for the Eddington-limited 
growth models with $z_{\rm i}=8$ (or $t_{\rm sup}=10^{8}\,{\rm yr}$) in 
Fig. 9,  and $z_{\rm i}=25$ (or $t_{\rm sup}=7\times 10^{8}\,{\rm yr}$) in 
Fig. 10. 
Figure 9 (top panel) shows that the peak AGN luminosity is 
$\sim 10^{45}\,{\rm erg/s}$. 
This is one order magnitude smaller than the threshold luminosity of QSOs 
(i.e., $L_{\rm QSO,th}=10^{46}\,{\rm erg/s}$) and thus the QSO phase 
does not appear. This is because the BH growth is 
greatly suppressed (see bottom panel in Fig. 9 and discussion on $\S 3.2$). 
Thus, the Eddington-limited growth model for $z_{\rm i}=8$ cannot explain 
the observed distant luminous QSOs. 
Also, the nuclear starburst activity dominates the AGN activity, and the peak nuclear starburst luminosity is $\sim 10^{47}\,{\rm erg/s}$. Such an intense starburst would be observed in the infrared band because of the strong dust extinction. If this is the case, we may discover a hyper-luminous infrared galaxy 
without a bright AGN at $z \sim 7$. Moreover, we predict a strong AGN outflow 
with $\dot{M}_{\rm outflow}\approx 10M_{\odot}\,{\rm yr}^{-1}$ in the nuclear 
starburst dominated phase (bottom panel in Fig. 9). Here the total outflow mass is obtained by $M_{\rm outflow}\equiv \int_{0}^{t}\dot{M}_{\rm outflow}(t^{\prime})dt^{\prime}$.  

\vspace{5mm}
\epsfxsize=8cm 
\epsfbox{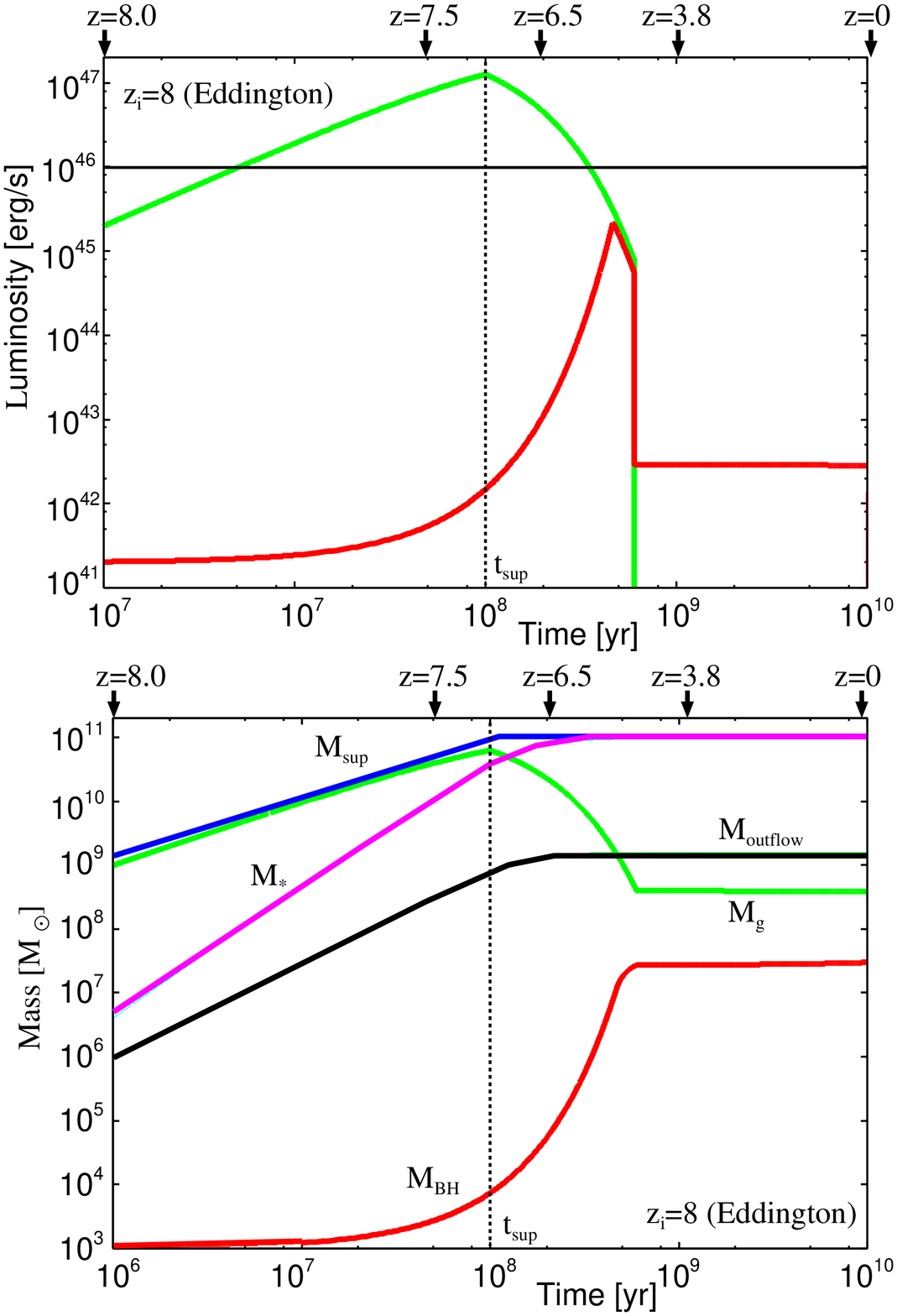}
\figcaption
{
(top panel): Same as Fig. 7 (a), but results for the Eddington-limitted 
growth model for $z_{\rm i}=8$ (or $t_{\rm sup}=10^{8}\,{\rm yr}$). 
(bottom panel): Same as Fig. 8 (a), but results for the Eddington-limitted 
growth model for $z_{\rm i}=8$ (or $t_{\rm sup}=10^{8}\,{\rm yr}$). 
The total outflow mass ($M_{\rm outflow}$) is newly added. 
In this case, QSO phase ($L_{\rm AGN} > L_{\rm QSO, th}$) does not appear. 
}
\vspace{2mm}

Figure 10 (top panel) shows that the peak AGN luminosity is comparable to $L_{\rm QSO, th}$ for the Eddington-limited growth model for $z_{\rm i}=25$.
This is because the final BH mass is comparable to that for the super-Eddington growth model (see Fig. 1 and the discussion on $\S 3.2$). 
Thus, this model can marginally reproduce the observable properties of 
QSOs at $z\sim 6$. However, it is hard to form more luminous distant QSOs 
at $z >8$ by the Eddington-limited growth as far as the radiative efficiency 
is $\epsilon_{\rm BH}\sim 0.1$. 
Thus, it is essential to discover the distant luminous QSOs 
by the future deep wide-field survey.
Figure 10 (top panel) also shows that the nuclear-starburst luminosity dominates the AGN luminosity in the proto-QSO phase (i.e., $L_{\rm SB}\gg L_{\rm AGN}$). This is different from the results of the super-Eddington growth model.
On the other hand, the evolution of $M_{\rm g}(t)$ and $M_{*}(t)$ are 
the same as the results of super-Eddington growth model (see Fig. 8 (a)). 
Thus, it is predicted that a stellar rich massive CND exists in the QSO phase 
and for the proto-QSO phase there exists a gas rich massive CNDs (see bottom panel in Fig. 10). 

\vspace{5mm}
\epsfxsize=8cm 
\epsfbox{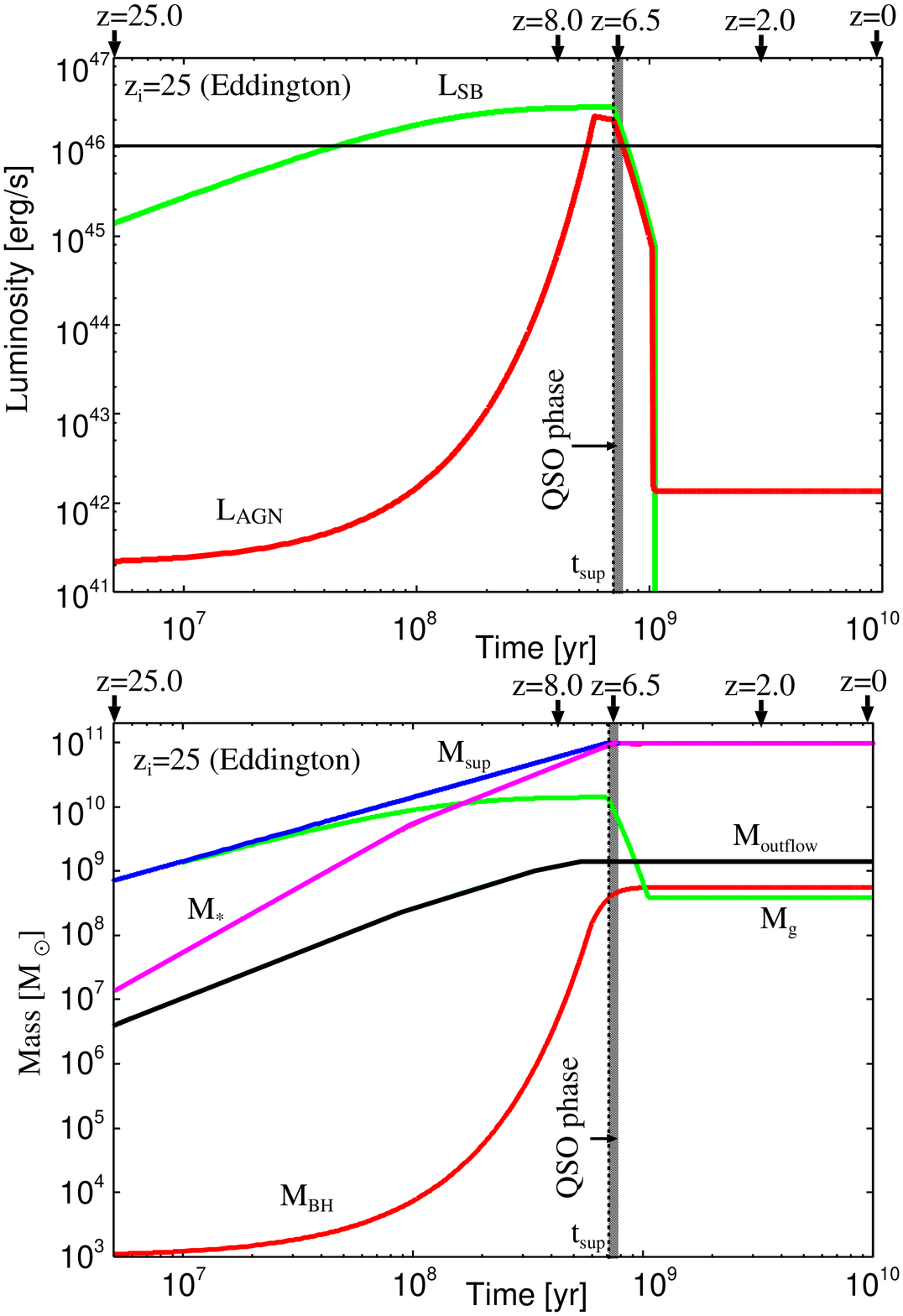}
\figcaption
{
(top panel): Same as Fig. 7 (a), but results for the Eddington-limitted 
growth model for $z_{\rm i}=25$ (or $t_{\rm sup}=7\times 10^{8}\,{\rm yr}$). 
(bottom panel): Same as Fig. 8 (a), but results for the Eddington-limitted 
growth model for $z_{\rm i}=25$ (or $t_{\rm sup}=7\times 10^{8}\,{\rm yr}$). 
The total outflow mass ($M_{\rm outflow}$) is newly added. 
The shaded region denotes the QSO phase. 
}
\vspace{2mm}

\section{Application to the distant QSO SDSS J1148+5251 at $z=6.42$}
We here compare our predictions ($\S 4$) with a distant QSO, the well known 
and examined SDSS J1148$+$5251 at $z=6.42$. 
We summarize the properties of this QSO elucidated by using 
multi-wavelengths. The bolometric QSO luminosity is $L_{\rm AGN}\sim 
10^{47}\,{\rm erg}\,{\rm s}^{-1}$ by near infrared observations 
(e.g., Willott et al. 2003; 
Barth et al. 2003). The radio and sub-mm observations suggest 
vigorous star formation in $\sim$ kpc with $\sim 10^{3}M_{\odot}
\,{\rm yr}^{-1}$ (Bertoldi et al. 2003a; Carilli et al. 2004; 
Walter et al. 2009). 
In addition, the molecular gas in the central $\sim$ 6 kpc is 
$\sim 5\times 10^{10}M_{\odot}$ by using carbon monoxide (CO) emission (Walter et al. 2003. 2004; Bertoldi et al. 2003b). However, we should note 
that the observed emission line might be sitting on top of a much broader line, because the low signal to noise ratio of the current data is not enough to 
determine the emission with accuracy (Narayanann et al. 2008). 

The observed high luminosity with $L_{\rm AGN}\sim10^{47}\,{\rm erg}\,{\rm s}^{-1}
$, could correspond to the late growth of SMBH with $z_{\rm i}=8$ (see Fig. 7 (b)) rather than the early growth of SMBHs with $z_{\rm i}=25$ (see Fig. 7 (a) and Fig. 10 (top panel)). 
If this is the case, the rapid and high mass accretion 
from $\sim$ kpc scale (i.e., $\dot{M}_{\rm sup}\approx 10^{3}M_{\odot}
\,{\rm yr}^{-1}$ and $M_{\rm sup}\approx 10^{11}M_{\odot}$) is required. 
Thus, the presence of massive CNDs with $M_{\rm g}=5\times 10^{10}M_{\odot}$ 
and a luminous nuclear starburst in the infrared band with $L_{\rm SB}\approx 
10^{47}\,{\rm erg}\,{\rm s}^{-1}$ are predicted. 
These values are larger than the case of the early growth of SMBHs with 
$z_{\rm i}=25$, i.e., $M_{\rm g}< 10^{10}M_{\odot}$ and $L_{\rm SB}
\approx 2\times 10^{46}\,{\rm erg}\,{\rm s}^{-1}$. 
The predicted CNDs can be observed through CO and hydrogen cyanide molecular 
emission using ALMA even at $z \sim 6$ because of $M_{\rm g}\approx 10^{10-11}M_{\odot}$ in the CND (e.g., Kawakatu et al. 2007). 
The PAH emission is a powerful indicator of the obscured starburst 
activity (e.g., Imanishi 2002). 
For 3.3 $\mu {\rm m}$ (the rest frame) PAH emission  from CNDs at $z=6$, 
$\sim 3\times 10^{-19}
\,{\rm W}\,{\rm m}^{-2}$ line flux is expected at $\sim 25\, \mu{\rm m}$ at 
the observed frame, assuming that the ratio of 3.3 $\mu {\rm m}$ PAH emission 
and the total infrared luminosity is $\approx 10^{-3}$ for pure starburst galaxies (e.g., Mouri et al. 1990; Imanishi 2002). 
The Space Infrared Telescope for Cosmology and Astrophysics 
(SPICA) and the James Webb Space Telescope (JWST) 
through polycyclic aromatic hydrocarbon (PAH) emission could also detect 
dusty starbursts in CNDs of this QSOs because the predicted value is above 
the line sensitivity limit (Swinyard et al. 2008). 
The present model can be verified by these future observations.

On the basis of the present results, we will constrain the host mass of the $z=6.42$ QSO SDSS J1148+5251. The mass accretion due to the tidal torque driven by the major and minor merger of galaxies is an essential process during the hierarchical formation of galaxies (e.g., Toomre \& Toomre 1972; Mihos \& Hernquist 1994, 1996; Saitoh \& Wada 2004; Saitoh et al. 2008). 
Saitoh \& Wada (2004) examined the evolution of stellar and gaseous cores 
on a sub-kpc scale during the formation of a spiral galaxy by hierarchical 
mergers, and found that the galactic core ($< 300$ pc) coevolves with the galactic dark matter (DM) halo of $\sim$10 kpc scale. The average mass ratio between the gas component in the core, $M_{\rm core}$ and the DM halo, $M_{\rm DM}$ is nearly constant, $M_{\rm core}/M_{\rm DM}\sim 0.04$. 
If we assume that the galactic core corresponds to the CNDs, i.e., 
$M_{\rm core}=M_{\rm sup}$, QSOs at $z >6$ would be formed in a massive 
DM halo of $M_{\rm DM} > 10^{12}M_{\odot}$ 
because of $M_{\rm sup} > 10^{10}M_{\odot}$. 
A massive halo of $\approx 10^{12-13}M_{\odot}$ was also suggested by 
Li et al. (2007) to explain the low space density of bright QSOs 
at $z\sim 6$ (e.g., Fan et al. 2001). 
Alternatively, Umemura (2001; hereafter U01) proposed another 
mass accretion process, i.e., the mass accretion onto the galactic center 
from a galactic scale ($\sim$ kpc) via the radiation drag 
(see also Kawakatu \& Umemura 2002). According to U01, the maximal mass 
accretion rate in the optically thick regime is 
given by $\dot{M}_{\rm drag}=L_{\rm host}/c^{2}\approx 
0.1M_{\odot}\,{\rm yr}^{-1}(L_{\rm host}/10^{12}L_{\odot})$, 
where $L_{\rm host}$ is the luminosity of host galaxies. 
To reproduce the rapid mass infall, e.g., $\dot{M}_{\rm sup}\approx 
10M_{\odot}\,{\rm yr}^{-1}$, the hosts of proto-QSOs 
must have strong star formation with 
$L_{\rm host}\approx 10^{14}L_{\odot}$. 
If this is the case, the stellar mass of the QSO host can be 
$M_{\rm host} \approx 10^{13}M_{\odot}$ because of $M_{\rm BH}\approx 
0.001M_{\rm host}$ (U01; Kawakatu, Umemura and Mori 2003). 
These discussions indicate that the SDSS J1148+5251 
form in massive dark matter halos and/or massive hosts 
(i.e., $M_{\rm DM}\approx 10^{12-13}M_{\odot}$ and/or  $M_{\rm host}
\approx 10^{13}M_{\odot}$), in order to build a SMBH with 
$>10^{9}M_{\odot}$ only by the gas accretion. 
This point should be tested by the future deep wide-field AGN survey.

Moreover, we discuss the nature of the progenitor of SDSS J1148+5251. 
As seen in Figure 7 (b), the AGN luminosity is still bright with 
$L_{\rm AGN}\approx 10^{47}\,{\rm erg}\,{\rm s}^{-1}$ at $z\sim 7.5$. 
It is hard to observe the AGN feature in the optical and X-ray observations 
because the AGN could be obscured by the dusty CND and dusty hosts 
(see Fig. 6). 
We predict that the progenitor can be super-Eddington objects (see Fig.7 (b)) 
and observed as high-$z$ ULIRGs such as SMGs 
with $L_{\rm IR}\approx 10^{47}\,{\rm erg}\, {\rm s}^{-1}$ because of 
$L_{\rm AGN}\approx L_{\rm SB}$. 
It is also predicted that the progenitor has a {\it gas rich} CND with 
the mass ratio $M_{\rm g}/M_{\rm dyn}\approx 1$. 
Such a gas rich CND with $M_{\rm g}\sim 10^{10-11}M_{\odot}$ can be 
detected by ALMA.

\section{Summary}
Based on the physical model of supermassive black hole (SMBH)
growth via gas accretion in $\sim $100 pc scale circumnuclear disks (CNDs) 
proposed by Kawakatu and Wada (2008), we investigate the formation of high-$z$ ($z > 6$) QSOs whose BH masses are $ M_{\rm BH}> 10^{9}M_{\odot}$. 
We show the necessary conditions to form QSOs at $z > 6$, in terms of 
the total gas mass accreted  from host galaxies, $M_{\rm sup}$ 
and star formation efficiency in the CND, $C_{*}=\dot{M}_{*}/M_{\rm g}$ 
where $\dot{M}_{*}$ and $M_{\rm g}$ are the star formation rate and 
the gas mass, respectively. 
Our main conclusions are summarized as follows. 

\begin{enumerate}
\item 
The required conditions for the formation of QSOs at $z >6$ are 
as follows;
(i) Rapid and large mass supply with $M_{\rm sup} > 10^{10}M_{\odot}$ within 1 Gyr from hosts to CNDs, because the final BH mass is only 
$1-10\%$ of the total mass supplied from hosts. 
The fraction is determined by the efficiency with which SN energy is 
transferred to the cold gas, and the efficiency of angular momentum transfer 
due to turbulent viscosity. (ii) High star formation 
efficiency, i.e., $10^{-8}\,{\rm yr}^{-1}< C_{*} < 10^{-6}\,{\rm yr}^{-1}$ is 
comparable to high-$z$ starburst galaxies such as SMGs. 
(iii) Rapid mass inflow from hosts with a period of the mass-supply, 
$t_{\rm sup}< 5\times 10^{9}\,{\rm yr}$.

\item 
We find that if the BH growth is limited by the Eddington accretion, 
the final BH mass is greatly suppressed when the period of mass-supply from 
hosts, $t_{\rm sup}$, is shorter than the Eddington timescale. 
Thus, the super-Eddington growth is required for the QSO formation, 
while $t_{\rm sup}$, which is determined by the efficiency of 
angular momentum transfer, is shorter than $\sim 10^{8}\,{\rm yr}$. 

\item
On the basis of these results, we predict the observable properties and 
the evolution of QSOs at $z >6$ as follows; 
(i) The evolution of the QSO luminosity strongly depends on the redshift $z_{\rm i}$ at which the accretion onto a seed BH got started. In other words, 
the bright QSOs given $z$ (e.g., $z=6$) favor the late growth of SMBHs (i.e., $z_{\rm i}\approx 10$) rather than the early growth (i.e., $z_{\rm i}\approx 30$) because the timescale of gas supplied from hosts is shorter than that of the 
star formation in the CND. 
(ii) In the QSO phase, after which the gas supply stopped and when 
the AGN luminosity is above the threshold one, $L_{\rm AGN, th}$
 (e.g., $L_{\rm AGN, th}=10^{46}\,{\rm erg}\, {\rm s}^{-1}$), 
 there exists a stellar rich massive CND with $M_{\rm g}/M_{\rm dyn}
\approx 0.1$ where $M_{\rm dyn}$ is the dynamical mass of the CND 
plus BH system.
On the other hand, in the proto-QSO phase (the early phase of a growing BH), 
the proto-QSO has a gas rich massive CNDs with $M_{\rm g}/M_{\rm dyn}\approx 1$.

\item
Our theoretical model predicts that the distant QSO SDSS J1148+5251 
discovered at $z=6.42$ 
favors the model of late growth of SMBH with $z_{\rm i}=8$ 
rather than early growth with $z_{\rm i}=25$. 
If this is the case, the QSO  harbors a massive CNDs with $M_{\rm g}=
5\times 10^{10}M_{\odot}$ and the luminous nuclear starburst in the infrared band with $L_{\rm SB}\approx 10^{47}\,{\rm erg}\,{\rm s}^{-1}$. 
These predictions can be checked by the future observations (e.g., ALMA, 
SPICA, and JWST), in order to judge whether our model is reasonable for 
the formation of SMBHs. In addition, we predict that SDSS J1148+5251 
have been evolved in a massive dark matter halo and/or a massive host 
(i.e., $M_{\rm DM}\approx 10^{12-13}M_{\odot}$ and/or 
$M_{\rm host}\approx 10^{13}M_{\odot}$), in order to build a SMBH with 
$>10^{9}M_{\odot}$ only by gas accretion. 
The progenitor of SDSS J1148+5251 can be super-Eddington objects and 
observed as the high-$z$ ULIRGs such as SMGs with the infrared luminosity 
$L_{\rm IR}\sim 10^{47}\,{\rm erg}\,{\rm s}^{-1}$. 
The progenitor has a {\it gas rich} CND with the mass ratio 
$M_{\rm g}/M_{\rm dyn}\approx 1$ and $M_{\rm g}\geq 10^{9}M_{\odot}$, 
which will be observed by ALMA. 

\end{enumerate}

\acknowledgments 
We appreciate many valuable suggestions and comments of anonymous referees. 
We thank A. Ferrara, F. Takahara and M. Umemura for useful comments 
and discussions. We thank T. Elizabeth for carefully reading the manuscript.
NK is financially supported by the Japan Society for 
the Promotion of Science (JSPS) through the JSPS Research Fellowship for 
Young Scientists.


\end{document}